\documentclass[%
 reprint,
 amsmath,amssymb,
 aps,
 prd,
 superscriptaddress,
 floatfix,
 longbibliography,
 noeprint
]{revtex4-2}
\usepackage{graphicx}
\usepackage{placeins}
\usepackage{subfigure}
\usepackage{bm}
\usepackage{multirow}
\usepackage{url}
\usepackage{hyperref}
\hypersetup{
colorlinks=true,
linkcolor=blue,
citecolor=blue,
}
\usepackage{makecell}
\usepackage{float}
\usepackage{tabto}
\usepackage{orcidlink}
\usepackage{tikz}
\newcommand\Tdiag[4]{%
    \multicolumn{1}{p{#2}}{\hskip-\tabcolsep
    \begin{tikzpicture}[%
                baseline={(0,-.25\baselineskip)},
                every node/.style={outer sep=0pt,inner sep=#1}]
    \node[minimum width={#2+1\tabcolsep-\pgflinewidth},
        minimum height=2\baselineskip-\pgflinewidth+\extrarowheight,
        use as bounding box] (box) {};
    \draw[line cap=round] (box.north west) -- (box.south east);
    \node[anchor=south west,text width=.75*#2,align=left] at (box.south west) {#3};
    \node[anchor=north east,text width=.75*#2,align=right] at (box.north east) {#4};
\end{tikzpicture}\hskip-\tabcolsep}}

\newcommand\dbline{\noalign{\vskip 0.10truecm\hrule\vskip 0.05truecm\hrule\vskip 0.10truecm}}
\newcommand\sgline{\noalign{\vskip 0.10truecm\hrule\vskip 0.10truecm}}
\newcommand\clineskip{\noalign{\vskip 0.10truecm}}
\usepackage{ulem}
\begin{document}

\begin{flushright}{\large PITT-PACC-2313-v2}\end{flushright}
\vspace{10mm}
\title{
Illuminating all-hadronic final states with a photon:
Exotic decays of the Higgs boson to four bottom quarks in vector boson fusion plus gamma at hadron colliders
}

\author{Stephen T. Roche\,\orcidlink{0000-0002-3878-5873}}
\email{stephen.roche@health.slu.edu}
\affiliation{School of Medicine, Saint Louis University, St.\ Louis MO, 63103, USA}
\affiliation{Department of Physics and Astronomy, University of Pittsburgh, Pittsburgh PA, 15260, USA}

\author{Benjamin T. Carlson\,\orcidlink{0000-0002-7550-7821}}\thanks{Corresponding author}
\email{bcarlson@westmont.edu}
\affiliation{Department of Physics and Engineering, Westmont College, Santa Barbara CA, 93108, USA}
\affiliation{Department of Physics and Astronomy, University of Pittsburgh, Pittsburgh PA, 15260, USA}

\author{Christopher R. Hayes\,\orcidlink{0000-0002-0298-0351}}
\email{hayeschr@umich.edu}
\affiliation{Department of Physics, University of Michigan, Ann Arbor MI, 48109, USA}

\author{Tae Min Hong\,\orcidlink{0000-0001-7834-328X}}
\email{tmhong@pitt.edu}
\affiliation{Department of Physics and Astronomy, University of Pittsburgh, Pittsburgh PA, 15260, USA}

\date{\today}

\begin{abstract}
\noindent We investigate the potential to detect Higgs boson decays to four bottom quarks through a pair of pseudoscalars, a final state that is predicted by many theories beyond the Standard Model. For the first time, the signal sensitivity is evaluated for the final state using the vector boson fusion (VBF) production with and without an associated photon, for the Higgs at $m_H=125\,\textrm{GeV}$, at hadron colliders. The signal significance is $0.5\sigma$ to $6\sigma$, depending on the pseudoscalar mass $m_a$, when setting the the Higgs decay branching ratio to unity, using an integrated luminosity of $150\,\textrm{fb}^{-1}$ at $\sqrt{s}=13\,\textrm{TeV}$. This corresponds to an upper limit of $0.3$, on the Higgs branching ratio to four bottom quarks, with a non-observation of the decay. We also consider several variations of selection requirements---input variables for the VBF tagging and the kinematic variables for the photon---that could help guide the design of new triggers for the Run-3 period of the LHC and for the HL-LHC. 
\end{abstract}

\maketitle

\section{Introduction}

After the discovery of the Higgs boson at $m_H=125\,\textrm{GeV}$ by the ATLAS and CMS collaborations in 2012 \cite{Aad:2012tfa,Chatrchyan:2012ufa}, the question of whether the observed particle shows any deviation from the Standard Model (SM) has been vigorously pursued. So far, the experimental observation of several decay and production properties have confirmed the hierarchy of Yukawa couplings \cite{Sirunyan:2018kst,Aad:2020jym,Aaboud:2018urx,Sirunyan:2018hoz,Khachatryan:2016vau,Aaboud:2018pen,Sirunyan:2017khh,ATLAS:2020fzp,ATLAS:2022ooq} and further studies of the CP properties are consistent with the SM \cite{ATLAS:2021pkb,ATLAS:2020ior,ATLAS:2020evk,ATLAS:2022tan,CMS:2021sdq,CMS:2022dbt}. The combined fits of these individual measurements constrain the decay width of the Higgs boson and put a constraint on the undetected non-SM fraction of all decays to be less than $0.12$ (ATLAS~\cite{ATLAS:2022vkf}) or $0.16$ (CMS~\cite{CMS:2022dwd}). These constraints do not include the undetectable decays as inputs to the combination. However, these fits make assumptions, such as the range of allowed couplings to vector bosons, that may not be involved in searches for specific final states predicted by theories beyond the Standard Model (BSM). Therefore, direct searches, such as those proposed in this paper, provide complementary information.

Due to the relatively narrow width of the SM Higgs boson $\Gamma_{H}\approx 4\,\textrm{MeV}$ \cite{Curtin:2013fra}, a small coupling to a new light state could lead to a branching ratio that is large enough to be observed at the Large Hadron Collider (LHC) while still evading current experimental constraints. A wide range of models in which the Higgs boson decays ``exotically'' to light scalars or pseudoscalars have been proposed \cite{SHROCK1982250,Dobrescu:2000yn,Ellwanger:2003jt,Ellwanger:2005uu,Stelzer:2006sp,Cheung:2007sva,Carena:2007jk,Chang:2008cw,Cao:2013gba}. In these models, such a particle is called $a$ and typically decays to SM particles. Following the Yukawa ordered coupling patterns of the Higgs boson, the branching ratio of $a$ to fermions scales with the mass of the fermion, so the decay $a\rightarrow b \bar{b}$ is the largest. For the kinematically accessible cases, the Higgs boson decays into two $a$ particles leading to the overall final state of $H\rightarrow aa \rightarrow b\bar{b}b\bar{b}$, which we write as $H_{4b}$. We consider the models in which the $a$ pseudoscalar decays promptly. In other models not examined in this paper, the $a$ particle can have a significant lifetime (see Ref.~\cite{Cepeda:2021rql} for a recent review). 

However, such exotic decays of the Higgs boson \cite{Cepeda:2021rql} present an experimental challenge at the LHC, as all-hadronic final states are often difficult to separate from the large background rate from QCD production of multiple hadronic jets (multijet). A variety of proposals exist to trigger on the lepton from the $W$ or $Z$ decay in the associated production of the Higgs boson for $H_{4b}$ \cite{Kaplan:2011vf,Cheung:2007sva,Carena:2007jk}, on $b$ physics triggers for $H_{4b}$ at LHCb \cite{Kaplan:2009qt}, and on the vector boson fusion (VBF) production mode for $H\rightarrow aa\rightarrow b\bar{b}\tau\tau$ \cite{Bomark:2015fga,Ellwanger_2005,Ellwanger:2003jt,dawson2008higgs}. The VBF production mode in particular has been shown to be of importance for $a$ particles with significant lifetime ~\cite{Curtin:2015fna}. For prompt signatures, VBF has been proposed for the $H\rightarrow 4\tau$ \cite{Adhikary:2022jfp} and $H \rightarrow \gamma\gamma jj$~\cite{Martin:2007dx} final states, the latter of which inspired an ATLAS search for $H \rightarrow \gamma\gamma jj$ final state~\cite{ATLAS:2018jnf}.

The ATLAS search using the $W$ or $Z$ associated production channel resulted in an upper limit on the branching ratio of $\mathcal{B}(H_{4b})$ for $m_a=60\,\textrm{GeV}$ at approximately $0.6$ with an expected value of $0.4$ at $95\%$ confidence level \cite{ATLAS:2018pvw,ATLAS:2020ahi}. No searches have yet been attempted for $H_{4b}$ in the VBF production channel, the topic of this paper, although promising searches for di-Higgs, $HH\rightarrow 4b$, exist \cite{ATLAS:2020jgy}.

Additional searches for exotic Higgs decays with cleaner final states, such as $H\rightarrow aa\rightarrow b\bar{b}\mu\mu$ allow for improved sensitivity \cite{ATLAS:2021hbr,CMS:2018nsh,CMS-PAS-HIG-22-007}.
However, some models---such as type-I two-Higgs doublet models called 2HDMs---prefer final states with $b$ quarks with relatively low branching ratios to muons~\cite{Haisch:2018kqx}, so $H_{4b}$ remains an important benchmark at the LHC. The model-specific branching ratio to the $4b$ final state is the product of the branching ratio of Higgs to $aa$ and $a\rightarrow bb$, $\mathcal{B} (H\rightarrow aa) \cdot \mathcal{B} (a\rightarrow b\bar{b})^2 $.
To simplify the presentation and for comparison with existing literature, we present results assuming $\mathcal{B}(H_{4b}$) = 1. 

Beyond the LHC, $H_{4b}$ is often suggested as a benchmark for searches at future colliders---such as the ILC~\cite{Baer:2013cma}, a proposed electron-positron collider, and the LHeC~\cite{LHeC:2020van, Liu:2016ahc}, an upgrade to the LHC to allow for $ep$ collisions.
A future electron-positron collider provides the cleanest final state and the strongest constraint, with a sensitivity of $\mathcal{B}(H_{4b}$) as low as $3\times10^{-4}$~\cite{Liu:2016zki,Carena:2022yvx}.
It is also reported in Ref.~\cite{Liu:2016zki} that the High Luminosity LHC (HL-LHC) will probe $\mathcal{B}(H_{4b}$) to around $0.2$.
Given the importance of $H_{4b}$ as a benchmark, the LHC experiments should consider various methods to target this final state during Run-3 as well as later at the HL-LHC.
In particular, trigger strategies should be considered that maximize the potential of this channel.  

\begin{figure}[b!]
\centering
    \includegraphics[width=\columnwidth]{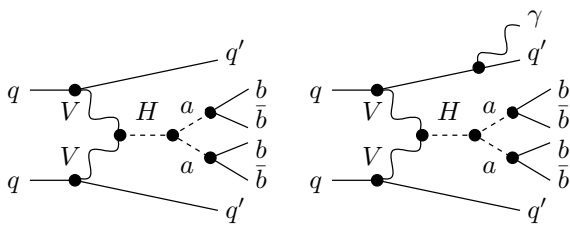}
    \caption{
    VBF production channel without a photon (left) and  with a photon (right) followed by the Higgs decay. In both cases, the Higgs boson decays to two pseudoscalars to four bottom quarks. Representative Feynman diagrams are shown and that the photon on the right can radiate from any of the charged particles, including from the vector bosons.
    \label{fig:feynman}
    }
\end{figure}

As the LHC experiments have explored $W$ or $Z$ associated Higgs production, we investigate the potential of the hadronic VBF final state, with and without a photon.
As previously mentioned, the multijet background is expected to dominate the VBF signal for $H_{4b}$. One proposed handle for the background is to require a photon produced in association with the VBF Higgs production \cite{Gabrielli:2007wf,Gabrielli:2016mdd,Asner:2010mj,Carlson:2021tes}. Figure~\ref{fig:feynman} shows a Feynman diagram of the VBF signal production channel without a photon (VBF$_{0\gamma}$) and with a photon (VBF$_{1\gamma}$). In addition to reducing the background, the photon also provides a handle for implementing a more efficient level-1 (L1) trigger, increasing the overall acceptance for the final state. The ATLAS experiment, for example, has already successfully utilized the photon in the VBF production channel in the search for the SM decay $H\rightarrow b\bar{b}$ \cite{Aaboud:2018gay,ATLAS:2020cvh} and the search for the BSM-enhanced invisible $H$ decay \cite{ATLAS:2021pdg}.
We show later that a similar photon strategy would be beneficial for targeting $H_{4b}$, and all production modes taken together will lead to a comprehensive search strategy, such as utilized in other exotic decays like Higgs to invisible~\cite{ATLAS:2023tkt,CMS:2023sdw}.

The relatively recent development of implementing machine learning (ML) methods for classification and regression, such as hls4ml's implementation of neural networks \cite{Duarte:2018ite,Loncar:2020hqp,Aarrestad:2021zos,Khoda:2022dwz} and boosted decision trees (BDT) by hls4ml/conifer \cite{Summers:2020xiy} and \textsc{fwXmachina} \cite{Hong:2021snb,Carlson:2022dgb}, on custom electronics boards with field programmable gate arrays (FPGA) has opened up a new era of possibility for ML-based L1 triggers. 
These tools have demonstrated the ability to perform ML classification of VBF production based on jets within 20 ns~\cite{Hong:2021snb}, which is well below the latency and resource constraints of L1 triggers at the LHC experiments~\cite{Achenbach:2008zzb,CMS:2020cmk}.
Therefore, we assume that the BDT taggers proposed in this paper are implementable at L1 using the above-mentioned tools.

In this paper, we present the first detailed estimate of the sensitivity of $H_{4b}$ using a BDT-based VBF tagger that could be implemented at the ATLAS or CMS L1 trigger. We also compare the $H_{4b}$ sensitivity of the BDT-based tagger to cut-based triggers inspired by triggers already implemented by the ATLAS and CMS experiments \cite{CMS:2020cmk,ATL-DAQ-PUB-2016-001,ATL-DAQ-PUB-2017-001,ATL-DAQ-PUB-2018-002,ATL-DAQ-PUB-2019-001}.

The paper is organized as follows. The next subsection describes the overall analysis strategy. Section~\ref{sec:data} describes the data sample, the event generation and simulation as well as the  detector setup and reconstruction techniques. Section~\ref{sec:selections} describes the data analysis, introducing a VBF tagger and Higgs taggers. Section~\ref{sec:results} presents the results, an estimate of the sensitivity for $L = 150\,\textrm{fb}^{-1}$ of integrated luminosity and a projection for the HL-LHC. Various trigger scenarios and methods for improving the signal sensitivity are also discussed. Section~\ref{sec:conclusions} concludes with an outlook for future searches.

\subsection{Analysis Strategy}

The strategy for separating signal from the SM background is as follows. First, a VBF tagger is used to target the VBF production mode of the Higgs boson. The sample is separated into two channels, VBF$_{0\gamma}$ and VBF$_{1\gamma}$ that contain events without and with a photon, respectively.

\begin{figure}[tb!]
    \centering
    \includegraphics[width=0.9\columnwidth]{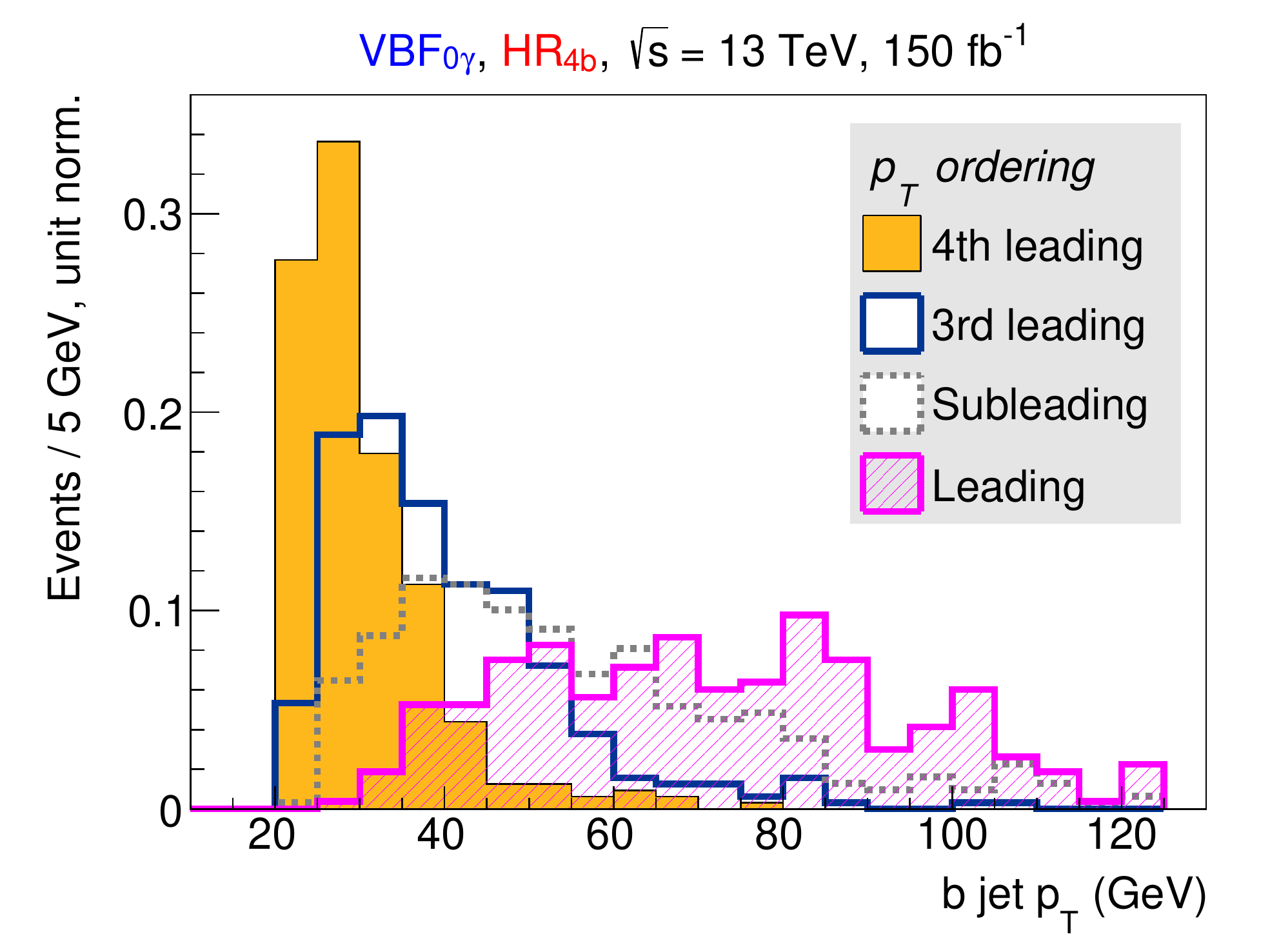}
    \caption{
    The $p_\textrm{T}$ distribution of each of the four $b$ jets from the $H_{4b}$ signal sample with $m_{a} = 50\,\textrm{GeV}$. These jets, as well as the VBF jets, are required to have $p_\textrm{T}>20\,\textrm{GeV}$ at reconstruction in Delphes. The plots are normalized to unity.
    \label{fig:bpt}
    }
\end{figure}

For the Higgs reconstruction (HR), two channels are considered corresponding to the number of reconstructed $b$ quark jets, HR$_{4b}$ and HR$_{3b}$. The former channel has six total reconstructed jets: two from the production and four from the Higgs decay. The HR$_{4b}$ channel has limitations, however, because the lowest-$p_\textrm{T}$ $b$ jet is generally below $40\,\textrm{GeV}$ and may evade identification. Figure~\ref{fig:bpt} shows the $p_\textrm{T}$ distribution for the VBF$_{0j}$ signal sample for $m_a=50\,\textrm{GeV}$ with the jet requirement of $p_{\textrm{T}} > 20\,\textrm{GeV}$. This selection requirement leads to a large fraction of events with only three reconstructed jets from the $H_{4b}$ decay. The latter $3b$ channel catches scenarios in which one of the $H_{4b}$ jets is not reconstructed. The loss of a jet can occur either due to merging with a nearby jet or if the jet fails one of the reconstruction requirements, which can cause it to fail the $b$ tagging selection, or a combination of both.
Examining the signal for $m_{a}=35\,\textrm{GeV}$, approximately $20\%$ of events had three b-flavored jets, while only approximately $3\%$ had 4 b-flavored jets. The fraction is lower for the $m_{a}=10\,\textrm{GeV}$ sample, for which $1.5\%$ of events had three b-flavored jets and $0.1\%$ of events had 4 b-flavored jets.
Therefore, we consider HR$_{3b}$ in which there are five total reconstructed jets.  A dedicated BDT-based tagger is used for each HR.

The combinations of the Higgs production and decay reconstruction gives rise to four analysis channels, which are given in Table~\ref{tab:channels}.  All taggers are described later in Sec.~\ref{sec:selections}.

\begin{table}[tb!]
\caption{
    The four analysis channels considered in this study. $O_\textrm{VBF}$ represents the VBF tagger. $O_{4b/3b}^{m_a}$ represents the Higgs taggers for the $4b$ and $3b$ Higgs reconstruction (HR) channels, respectively, for a given $m_a$. Different values of $m_a$ change the kinematic distributions of the final state, thus require different BDT.
    \label{tab:channels}
}
\begin{tabular}{
p{0.180\textwidth}
p{0.105\textwidth}
p{0.155\textwidth}
}
\dbline
\Tdiag{.2em}{0.180\textwidth}{Reconstruction}{Production}
& VBF$_{0\gamma}$
& VBF$_{1\gamma}$ channel \\
\sgline
HR$_{4b}$ channel & $O_{4b}^{m_a}$, $O_\textrm{VBF}$
                  & $O_{4b}^{m_a}$, $O_\textrm{VBF}$, req.\ $\gamma$ \\
HR$_{3b}$ channel & $O_{3b}^{m_a}$, $O_\textrm{VBF}$
                  & $O_{3b}^{m_a}$, $O_\textrm{VBF}$, req.\ $\gamma$ \\
\dbline
\end{tabular}
\end{table}

\section{Simulated data samples}
\label{sec:data}

Approximately $1.5$ billion events are generated for this study using Monte Carlo simulation. The event simulation (Sec.~\ref{sec:simulation}) is described followed by the event reconstruction (Sec.~\ref{sec:reco}). A subsample of about $150$ million events---those that pass a loose set of requirements on the number of reconstructed jets specified in Sec.~\ref{sec:reco}---can be found online \cite{dataset}.

The data sample collected during in the Run-3 data taking period is expected to be around $150\,\textrm{fb}^{-1}$~\cite{Fartoukh:2790409}. Our main study (Sec.~\ref{sec:sensitivity}) is optimized for this integrated luminosity. The study is extended to HL-LHC (Sec.~\ref{sec:hl-lhc}) by scaling the luminosity to $3\,\textrm{ab}^{-1}$.

\subsection{Event simulation}
\label{sec:simulation}

The signal and background samples considered are listed in Table~\ref{table:xs}. All samples are generated at leading order (LO) with MadGraph5 v2.7.3~\cite{Alwall:2014hca} with the Standard Model configuration. The generated event is then passed to Pythia v8.306 for the parton shower and hadronization process~\cite{Sjostrand:2014zea,Bierlich:2022pfr} using the ATLAS AZ Tune 17~\cite{ATLAS:2014alx}. All events are generated with a proton-proton center-of-mass energy of $\sqrt{s}=13\,\textrm{TeV}$, which was the value for Run-2 operations. For the Run-3 operations that started in 2022, the LHC increased the center-of-mass energy to $13.6\,\textrm{TeV}$. A previous scaling study \cite{Carlson:2021tes} demonstrated a linear increase in cross section from $\sqrt{s}$ from $13$ to $13.6\,\textrm{TeV}$, and therefore this percent-level difference is not considered here.

The background processes are weighted according to the cross sections stated in Table~\ref{table:xs}. These cross sections are calculated by MadGraph5 at leading order during sample generation. Notationally, final states such as $b\bar{b} + jj$ involve requirements on the parton selections passed to MadGraph, which in this case correspond to two $b$-quark jets and two light-flavor jets $j$. The latter refers to gluons as well as quarks with a smaller mass than the bottom quark. For each sample, the following requirements are applied. Hadronic jets were produced with a minimum $p_\textrm{T}$ of $20\,\textrm{GeV}$ and an angular separation $\Delta R >  0.4$ between every jet, where $\Delta R = \sqrt{\Delta\phi^2 + \Delta y^2}$ is the distance in azimuthal angle $\phi$ and rapidity $y$. Photons are required to have a minimum $p_\textrm{T} > 10\,\textrm{GeV}$ and a maximum pseudorapidity $|\eta| < 2.5$. Photons are also required to be separated from any hadronic jets by $\Delta R > 0.4$ at the generator level. 

\begin{table}[bt!]
\caption{
    List of samples used and the corresponding cross sections. The $j$ refers to light-flavor hadronic jets.
    \label{table:xs}
}
\centering
{\small
\TabPositions{0.20\columnwidth}
\begin{tabular}{
    p{0.29\textwidth}
    p{0.10\textwidth}
    p{0.09\textwidth}
}
\dbline
Sample &
\multicolumn{2}{l}{Cross section [pb]} \\
\sgline 
Signal, $m_H=125\,\textrm{GeV}$                                \\
\quad VBF$_{0\gamma}$ production channel & $3.8$                & \cite{LHCxswg:2016ypw} \\
\quad VBF$_{1\gamma}$ production channel & $9.45\times 10^{-2}$ & \cite{ATLAS:2021pdg}   \\
Background for VBF$_{0\gamma}$                                 \\
\quad $b\bar{b}$                         & $2.83\times 10^6$   \\
\quad $b\bar{b} + j$                     & $9.86\times 10^5$   \\
\quad $b\bar{b} + jj$                    & $5.04\times 10^5$   \\
\quad $b\bar{b} + b\bar{b}$              & $1541$              \\
\quad $t\bar{t}$                         & $505$               \\
\quad $Z_{bb} + b\bar{b}$                & $10.4$              \\
Background for VBF$_{1\gamma}$                                 \\
\quad $b\bar{b}          $\tab$+\gamma$  & $738$               \\
\quad $b\bar{b} + j      $\tab$+\gamma$  & $732$               \\
\quad $b \bar{b} + jj    $\tab$+\gamma$  & $433$               \\
\quad $t\bar{t}          $\tab$+\gamma$  & $2.1$               \\
\quad $Z_{bb} + jj       $\tab$+\gamma$  & $1.8$               \\
\quad $Z_{bb} + b\bar{b} $\tab$+\gamma$  & $1.6\times 10^{-2}$ \\
\dbline
\end{tabular}
}
\end{table}

The signal samples are produced with the Higgs mass of $125\,\textrm{GeV}$ for the two production channels, VBF$_{0\gamma}$ and VBF$_{1\gamma}$.
The same jet and photon selections at generator-level, as described for the background processes, are applied.
The cross section of VBF$_{0\gamma}$ is normalized to the prediction from the LHC Higgs Working Group~\cite{LHCxswg:2016ypw}, computed at next-to-next-to-leading-order (NNLO) in QCD using proVBF~\cite{Dreyer:2018rfu} and includes corrections at next-to-leading-order (NLO) from electroweak and photon processes computed using HAWK~\cite{Denner:2014cla}.
The cross section of VBF$_{1\gamma}$ is taken from the Auxiliary Material of Ref.~\cite{ATLAS:2021pdg}, computed at NLO using MadGraph5\_aMC@NLO v2.6.2 with the parton shower computed using Herwig v7.1.3p1~\cite{Bellm:2015jjp,Bahr:2008pv}. 
There is no explicit photon veto for VBF$_{0\gamma}$ and there is no overlap with VBF$_{1\gamma}$ at the event generation level.
We verified that approximately $1\%$ of VBF$_{0\gamma}$ events contain a photon after showering and reconstruction.

After the Higgs production, the Higgs boson is decayed to $H_{4b}$ using Pythia v8.306 without any generator-level cuts. Four separate decays are considered corresponding to pseudoscalar masses $m_{a}{\,=\,} 10, 25,35$ and $50\,\textrm{GeV}$. In summary, four signal samples are made for the combination of the two production and four decay processes.

Ideally, processes with six or more partons would be simulated with many of these partons being $b$ quarks. However, it is computationally expensive to simulate such events with high parton multiplicity. In MadGraph each additional parton increases the simulation time by approximately an order of magnitude. For instance, the simulation time for events with more than four partons is $\mathcal{O}(1)$ hour \cite{Alwall:2011uj}. Therefore, it is not practical to generate some potentially relevant background samples such as $b\bar{b}b\bar{b}+jj+\gamma$ with a statistically robust sample size. Our solution to address the sample statistics is twofold. First, we truncate the number of partons required in the MadGraph generation at four total jets at generator level, such as $b\bar{b}+jj$, for the background processes and rely on Pythia to produce additional jets via parton shower. Second, we weight the events by the product of $b$ tagging probabilities for the non-VBF jets.

At this point, we introduce some notation to ease our discussion. The two jets with the highest invariant mass \cite{Aaboud:2018gay} are denoted as the ``VBF jet pair'' and are labeled with uppercase letters $JJ$. Individually, they are distinguished as $J_1$ and $J_2$, with the ordering to denote the leading and subleading $p_\textrm{T}$ jet, respectively. To validate this decision, a sample of $100$k VBF$_{0\gamma}$ signal events are investigated. In only $8\%$ of those events, one or both of the jets in the pair with the highest invariant mass included a $b$ quark and therefore arise from the Higgs decay rather than the VBF process. We conclude that for the signal process, choosing the jet pair with the highest invariant mass correctly selects those arising from VBF production in $92\%$ of events.

The remaining jets, four of them for HR$_{4b}$ and three of them for HR$_{3b}$, are candidate $b$ quarks with subscripts labeled according to their large-to-small sorted $p_\textrm{T}$ values as $b_1$, $b_2$, $b_3$, and $b_4$, with the last jet only considered for HR$_{4b}$.

In the generation of background processes, we limit MadGraph at four jets with additional parton shower jets from Pythia. After identifying the VBF jets $JJ$, the remaining $b_i$ jets are considered for their $b$ tagging probability, $\epsilon_b$. Instead of requiring that the remaining jets be $b$ tagged, we consider the product of their $b$ tagging probability as the event weight \cite{Nackenhorst:2015yjt,Connelly:2016nmt} using the $\epsilon_b$ function of the CMS Delphes card, derived from Ref.~\cite{CMS:2012feb}.

We validated our two-fold approach as follows. The truncation is validated using three samples---corresponding to $b\bar{b}+j$, $b\bar{b}+jj$, and $b\bar{b}+3j$ at parton level---of around 100k events each. As described above, any additional jets in the sample are acquired from the parton shower in Pythia. Using the $b$ event weight described above, we compare the number of events from each sample with four reconstructed $b$ jets. The difference with respect to the prediction for the number of events is found to be within $10\%$.

The $b$ event weight method is validated in a simulated multijet sample by comparing the number of events with four $b$ tags to the expected number from the weighting scheme. The difference is found to be within $1\%$.

\subsection{Event reconstruction}
\label{sec:reco}

The detector simulation is performed  using Delphes 3.5 and object reconstruction is done using Delphes-based algorithms~\cite{deFavereau:2013fsa,Ovyn:2009tx}. Detector parameters are selected to match the CMS detector as implemented using the CMS card without pileup. 

Hadronic jets with a $p_\textrm{T}$ of at least $20\,\textrm{GeV}$ are reconstructed with the FastJet program~\cite{Cacciari:2011ma} using the anti-$k_t$ algorithm with a radius parameter of $0.4$~\cite{Cacciari:2008gp}. In order to ensure that the CMS card is appropriately reconstructing jets, the $p_\textrm{T}$ resolution is evaluated using QCD multijet events and reconstructing generator-level anti-$k_t$ jets with the same parameters. If the nearest generator-level jet is within $\Delta R \leq 0.3$ of the reconstructed jet, then it is assumed that the matching reconstructed jet came from that generator-level jet. The jet energy resolution is found to be approximately $10\%$, consistent with the values reported by ATLAS~\cite{ATLAS:2020cli} and CMS~\cite{CMS-DP-2021-033}. 

The identification of jets containing $b$ hadrons ($b$ tagging) is performed by first identifying the flavor of the parton in the jet, then assigning the corresponding efficiency using the values given in Ref.~\cite{CMS:2012feb}. The $b$-tagging efficiency, $\varepsilon_b$, is a function of jet flavor and jet $p_\textrm{T}$ \cite{CMSBTag}.

The tagging efficiency expression reduces to a maximum efficiency of $70\%$ for jets with a bottom quark, $20\%$ for jets with a charm quark and $1\%$ for light-flavor jets. Improvements in $b$ tagging, such as those from ML-based algorithms are not included here~\cite{ATLAS:2022rkn,Bols:2020bkb}.

Photon candidates are reconstructed with a minimum $p_\textrm{T}$ of $10\,\textrm{GeV}$ and $|\eta| < 2.5$. Each photon is weighted using an efficiency factor that depends on its $p_\textrm{T}$ and $\eta$~\cite{delphesPhotonEfficiency}.

\section{Data analysis}
\label{sec:selections}

The analysis of the MC samples described in the previous section consists of three parts. The trigger assumptions are stated  (Sec.~\ref{sec:analysis}) and target only the VBF jets. The VBF production of the Higgs boson, with and without a photon, is selected using a single VBF tagger (Sec.~\ref{sec:vbf_tagger}) independent of the Higgs decay. The $H_{4b}$ decay is selected for the two reconstruction channels using Higgs taggers (Sec.~\ref{sec:higgs_tagger}) independent of the VBF production.

\subsection{Trigger}
\label{sec:analysis}

Multipurpose experiments, such as ATLAS~\cite{ATLAS:2012nks,Aaboud:2016leb} and CMS~\cite{CMS:2016ngn}, utilize two-level trigger systems. The first-level L1 trigger evaluates algorithms within a few microseconds in hardware \cite{Achenbach:2008zzb,ATLAS:2020osu,CMS:2020cmk}, while the second high level trigger (HLT) evaluates algorithms in software~\cite{Stelzer:2011zz} in $\mathcal{O}(1)$ second. In the current configuration of ATLAS and CMS, the L1 trigger system is limited in information and primarily utilizes only information from the muon and calorimeter sub-detectors.

For the HL-LHC data-taking, CMS plans to incorporate tracking into L1~\cite{Zabi:2020gjd}.  ATLAS upgrades are also planned~\cite{ATLAS:2013lic,ATLAS-TDR-029} to expand the capabilities of the calorimeter reconstruction for L1.  While these upgrades will undoubtedly improve the general capabilities of the trigger system, we focus on variations that are similar to the current trigger capabilities available in Run 2 and Run 3.

The trigger strategy followed in this paper, as well as the current stated strategy of ATLAS and CMS, is to use only the VBF jets. The VBF jets are generally higher in $p_\textrm{T}$ compared to the $b_i$ jets from the Higgs decay as was discussed in Sec.~\ref{sec:simulation}. Therefore, the Higgs decay products or the Higgs taggers are not considered for triggering.

The VBF tagger presented here makes favorable trigger assumptions, such as the presence of all reconstructed hadronic jets with the reconstruction thresholds $p_\textrm{T}>20\,\textrm{GeV}$ as inputs to $O_\textrm{VBF}$. In Sec.~\ref{sec:trigger}, we compare our BDT method to VBF taggers using various cut-based approaches for the trigger that drop some of these assumptions.

\subsection{VBF tagger}
\label{sec:vbf_tagger}

Hadronic jets arising from VBF Higgs production have a distinct kinematic signature. This signature includes a high dijet invariant mass, a large difference in $\eta$ between the jets, and a small separation in $\phi$ relative to the QCD multijet background. Using the pair of VBF jets $JJ$ defined in the previous section, we develop a BDT-based VBF tagger using $J_1$ and $J_2$.

TMVA~\cite{Hocker:2007ht} is used to train a forest of $200$ decision trees at a maximum depth of $6$ using the adaptive boost algorithm~\cite{adaboost}. The input variables to the BDT, listed below, are based on the two VBF jets $J_1$ and $J_2$.
\begin{itemize}
    \itemsep=0pt
    \item $m_{JJ}$, dijet invariant mass
    \item $p_{\textrm{T},J1}$, transverse momentum of leading VBF jet
    \item $p_{\textrm{T},J2}$, transverse momentum of subleading VBF jet
    \item $E_{J1}$, energy of the leading VBF jet \cite{Barger:1991ib}
    \item $E_{J2}$, energy of the subleading VBF jet \cite{Barger:1991ib}
    \item $|\Delta \phi_{JJ}|$, azimuthal separation \cite{Eboli:2000ze}
    \item $|\Delta \eta_{JJ}|$, pseudorapidity gap
\end{itemize}
The additional jets that are needed for the two HR channels are not used as inputs. Distributions of the input variables can be found in Ref.~\cite{Hong:2021snb}.

\begin{figure}[b!]
\includegraphics[width=0.90\columnwidth]{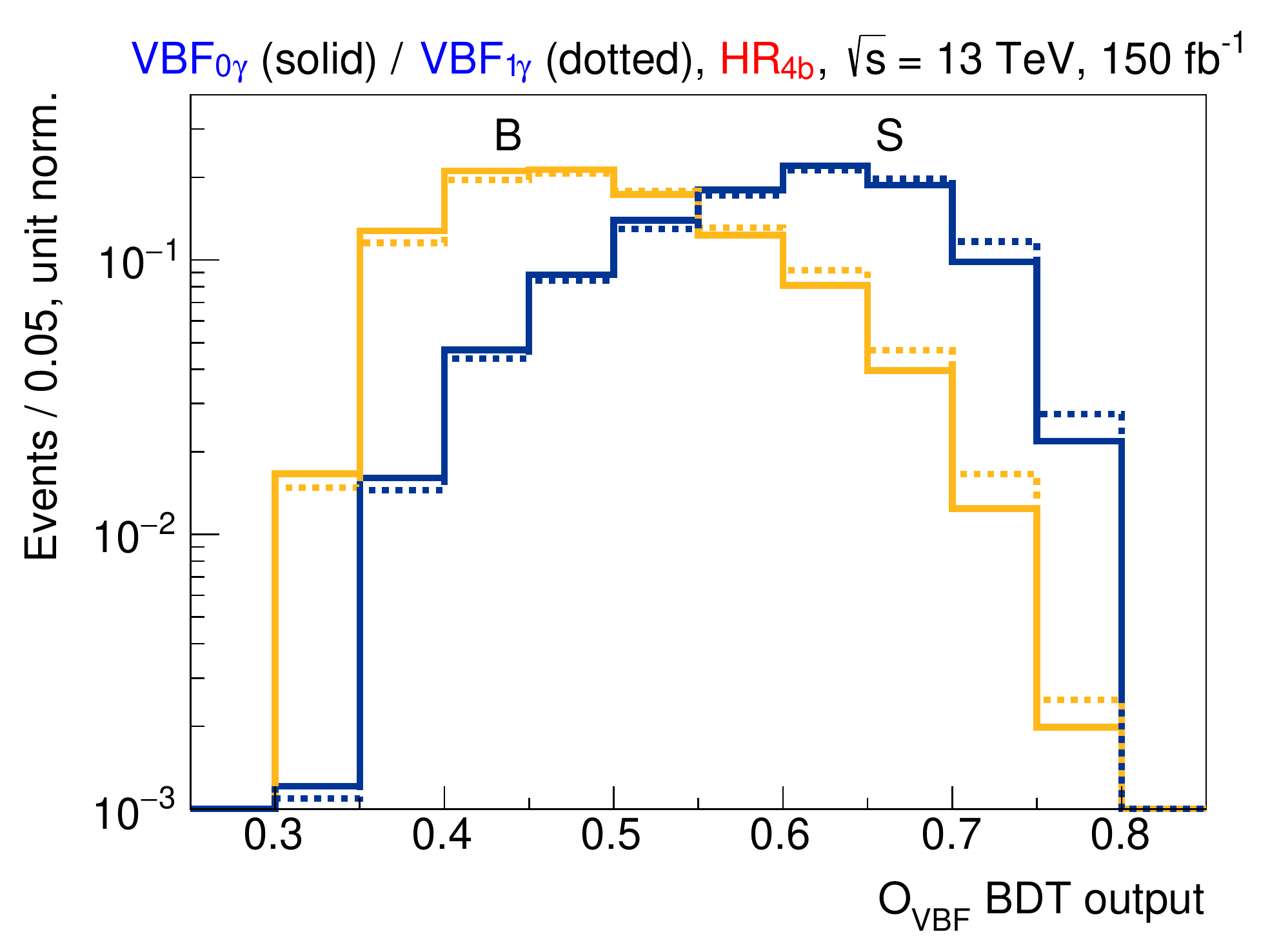}
\caption{
    VBF tagger score distribution. The score distributions are shown for the BDT using only the VBF jets. The unit-normalized distributions are shown for the $m_a=50\,\textrm{GeV}$ the signal ($S$, dark-colored lines) and SM background ($B$, light-colored lines). The VBF$_{0\gamma}$ (solid lines) and VBF$_{0\gamma}$ (dotted lines) show similar distributions for the $S$ and for the $B$.
    \label{fig:VBFtagger}
}
\end{figure}

The samples used for the BDT training are the signal and background samples for the VBF$_{0\gamma}$ channel, combining the HR$_{4b}$ and HR$_{3b}$ channels. The output score of the VBF BDT is called $\mathcal{O}_\textrm{VBF}$. Figure~\ref{fig:VBFtagger} shows the distributions for VBF$_{0\gamma}$ (solid lines).

We find no need to train a separate tagger for the VBF$_{1\gamma}$ channel. Figure \ref{fig:VBFtagger} shows that the signal-background separation for the BDT outputs evaluated on the VBF$_{1\gamma}$ sample without a photon requirement (dotted lines) is similar to that of VBF$_{0\gamma}$ (solid lines). The VBF$_{1\gamma}$ analysis channel requires a photon with $p_\textrm{T}{\,>\,}15\,\textrm{GeV}$ in addition to passing the VBF BDT requirements.

\subsection{Higgs taggers}
\label{sec:higgs_tagger}

A BDT is also developed to evaluate the non-VBF jets, $b_1$ to $b_4$ for HR$_{4b}$ and $b_1$ to $b_3$ for HR$_{3b}$. In combination with the two pseudoscalar mass scenarios, $m_a=25$ and $50\,\textrm{GeV}$, this results in four Higgs tagger BDTs with scores denoted as $O_{4b}^{50}$, $O_{3b}^{50}$, $O_{4b}^{25}$, and $O_{3b}^{25}$.

In the following subsections, the BDT training strategies for the HR$_{4b}$ and HR$_{3b}$ channels are described. The strategies are identical for the each of the $m_a$ masses with the only differences being the input signal samples.

\subsubsection*{HR\texorpdfstring{$_{4b}$}{4b} channel}

Four $b$ candidate jets are paired to find the two jet pairs associated with each $a\rightarrow b\bar{b}$ decay. The two jet pairs will have nearly the same invariant mass, $m_{bb}$, if they are decays of the $a$ particle. Therefore, the jet pairings are chosen such that the absolute difference in the sets of the invariant mass, $\Delta m_{bb} = |m(b_i,b_{i'}) - m(b_{ii},b_{ii'})|$, is minimized between two dijet pairs.

TMVA is configured using the same setup as the VBF BDT with the following three input variables:
\begin{itemize}
    \itemsep=0pt
    \item $m_{4b}$, the invariant mass of the four-jet system
    \item $\Delta m_{bb}$, the mass difference between the dijet pairs
    \item $m_{bb}^\textrm{avg}$, the average mass of the dijet pairs
\end{itemize}

\begin{figure*}[b!]
\centering
\includegraphics[width=0.90\textwidth]{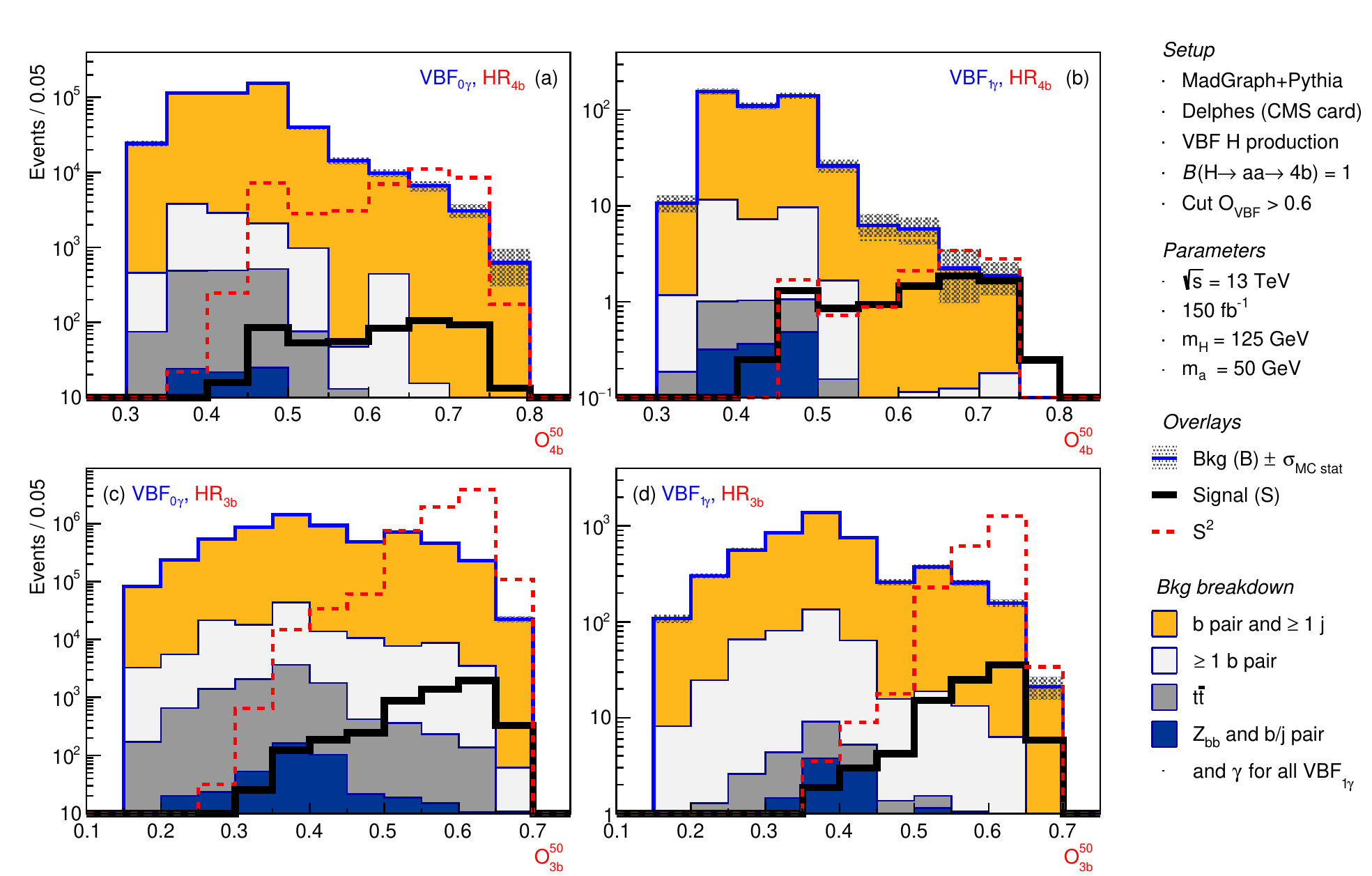}
\caption{
    Higgs tagger score distributions. The distribution $O_{4b}$ ($O_{3b}$) from the BDT for the two Higgs reconstruction channels HR$_{4b}$ (HR$_{3b}$) is given in the top row (bottom row). The signal ($S$) is for $m_a=50\,\textrm{GeV}$ assuming $\mathcal{B}(H_{4b}$) = 1 and the SM background ($B$) are shown for VBF$_{0\gamma}$ (VBF$_{1\gamma}$) in the left column (right column). Also shown is the squared event count for the signal ($S^2$, dotted line) so that $S/\sqrt{B}$ can be visualized. The statistical error on the MC statistics is shown as the shaded boxes around the background prediction. A selection on the VBF tagger, $O_\textrm{VBF}>0.6$, is applied.
    \label{fig:Htagger}
}
\end{figure*}

The BDT score distributions for $m_a=50\,\textrm{GeV}$ for the two VBF$_{0/1\gamma}$ channels are shown in the first row of Fig.~\ref{fig:Htagger}. The largest background contribution is from $b\bar{b}$ with one or more jets produced at LO, which is two orders of magnitude larger than the other backgrounds.

\subsubsection*{HR\texorpdfstring{$_{3b}$}{3b} channel}

When events are reconstructed with only three $b$ jets, different selections are used.

TMVA is configured as before. As it is not possible to pair the jets as was done for HR$_{4b}$, the following six input variables are used:
\begin{itemize}
    \itemsep=0pt
    \item $m_{3b}$, the invariant mass of the three jet system
    \item $m_{b1,b2}$, the invariant mass of the leading and subleading $b$ jets
    \item $m_{b1,b3}$, invariant mass of the leading and third-leading $b$ jets
    \item $m_{b2,b3}$, invariant mass of the non-leading $b$ jets
    \item $\Delta R_{bb}^\text{avg}$, average $\Delta R_{bb}$ of the three dijet pairs~\cite{ATLAS:2018pvw}
    \item $\Delta R_{bb}^\text{min}$, minimum $\Delta R_{bb}$ of the three dijet pairs~\cite{ATLAS:2018pvw}
\end{itemize}

The BDT score distributions for the $m_a=50\,\textrm{GeV}$ signal in the two VBF$_{0/1\gamma}$ channels are shown in the second row of Fig.~\ref{fig:Htagger}. A similar pattern of background contributions as HR$_{4b}$ is seen.

\section{Results}
\label{sec:results}

We present the expected sensitivity using the expected Run-3 data sample of $150\,\textrm{fb}^{-1}$ (Sec.~\ref{sec:sensitivity}), trigger proposals based on our study (Sec.~\ref{sec:trigger}), and HL-LHC projections for $3\,\textrm{ab}^{-1}$ (Sec.~\ref{sec:hl-lhc}).

\subsection{Expected sensitivity for LHC Run-3}
\label{sec:sensitivity}

The expected sensitivity to $H_{4b}$ is estimated using a counting experiment. As the backgrounds in a more complete experimental setting will likely rely on data-driven background estimation procedures for the multijet background, no attempt at an estimation of the statistical impact of the simulated sample size or of systematic uncertainties is undertaken. Though systematic uncertainties are expected to reduce the sensitivity, the result is expected to be statistically limited and this choice would not change the conclusions presented here.

Using the VBF tagger and Higgs taggers described in the previous section, we count the numbers of signal ($S$) and background events ($B$). An output score selection is chosen to maximize the sensitivity $\sigma$ quantified by $S/\sqrt{B}$, which is found to be greater than $0.6$ for $O_\textrm{VBF}$ and $O^{50/25}_{4b/3b}$. The distributions shown in Fig.~\ref{fig:Htagger} are after the VBF tagger selection on $O_\textrm{VBF}$, as well as the photon requirement for the VBF$_{1\gamma}$ channel.

\begin{figure*}[b!]
\centering
\includegraphics[width=0.990\textwidth]{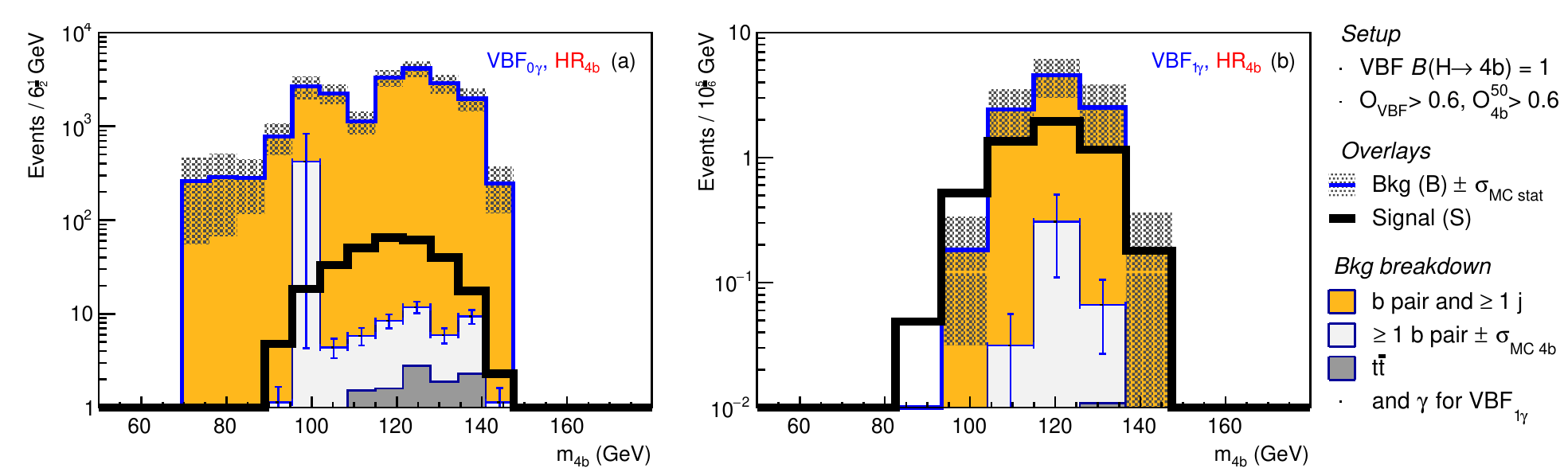}%
\caption{
    Four $b$ invariant mass. Distributions are shown for HR$_{4b}$ for the two production channels VBF$_{0\gamma}$ (left) and VBF$_{1\gamma}$ (right). The signal is shown for $m_a=50\,\textrm{GeV}$ and selections are made on the Higgs tagger $O_{4b}^{50}>0.6$ and the VBF tagger $O_\textrm{VBF}>0.6$. We refer to the caption of Fig.~\ref{fig:Htagger} for the technical aspects of the plots.
    \label{fig:m4b}
}
\end{figure*}

For HR$_{4b}$, the signal significance in VBF$_{0\gamma}$ after $O_{4b}^{50}$ is $2.1\sigma$, corresponding to a signal acceptance of $0.43$ and background rejection of $0.993$. The signal significance in VBF$_{1\gamma}$ after $O_{4b}^{50}$ is $1.7\sigma$, corresponding to a signal acceptance of $0.46$ and background rejection of $0.998$. (Acceptance is defined as $\varepsilon=N_\textrm{pass}/N_\textrm{total}$ while rejection is defined as $1-\varepsilon$.) The $m_{4b}$ distributions after selections on both the VBF and the Higgs tagger is shown in Fig.~\ref{fig:m4b}.

For HR$_{3b}$, the signal significance in VBF$_{0\gamma}$ after the $O_\textrm{3b}^{50}$ selection is $4.6\sigma$, corresponding to a signal acceptance of $0.62$ and background rejection of $0.97$. The signal significance in VBF$_{1\gamma}$ after the $O_\textrm{3b}^{50}$ selection is $3.1\sigma$, corresponding to a signal acceptance of $0.64$ and background rejection of $0.97$.

The above results for $m_a = 50\,\textrm{GeV}$, as well as the results for $m_a=10,25,35\,\textrm{GeV}$, are given in Table~\ref{tab:sigma}, assuming $\mathcal{B}(H_{4b})=1$ and an integrated luminosity of $150\,\textrm{fb}^{-1}$. A total sensitivity of between 0.5 and $6\sigma$, depending on $m_{a}$, is obtained by combining the individual channels.
The signal efficiency is greatly reduced for $m_{a} = 10\,\textrm{GeV}$ because the b-jets from the $a$-decay become highly collimated and fewer events pass the b-tagging selections.
These final-states with collimated jets could be searched for using alternative analysis strategies, such as those proposed in Ref.~\cite{Curtin:2014pda}.
In the absence of a clear signal, an upper limit on the branching ratio can be derived assuming the SM Higgs cross-section. The upper limit on the branching ratio of Higgs to four bottom quarks is estimated using $2/\sigma$, the approximate statistics-only sensitivity estimate at $95\%$ confidence level, for $m_a \ 50 \textrm{GeV}$ and is found to be $0.33$. 

\begin{table*}[btp!]
\caption{
    Event counts for the number of signal $S$ assuming $\mathcal{B}(H_{4b}$) = 1 and background $B$ (top half) and estimated sensitivity $\sigma$ (bottom half) using $150\,\textrm{fb}^{-1}$ of data. The analysis is performed  with a BDT VBF trigger and BDT Higgs tagger for the signal samples with $m_a = 50\,\textrm{GeV}$ (a) and $25\,\textrm{GeV}$ (b). In this table, $n_j$ refers to the number of hadronic jets of any flavor and $n_b$ the subset identified as $b$ quark jets.
\label{tab:sigma}
}
\centering
{
\subtable[Event counts]{
\begin{tabular}{
p{0.280\textwidth}
p{0.080\textwidth}
p{0.080\textwidth}
p{0.080\textwidth}
p{0.080\textwidth}
p{0.080\textwidth}
p{0.080\textwidth}
p{0.080\textwidth}
p{0.080\textwidth}
}
\dbline
& \multicolumn{4}{l}{VBF$_{0\gamma}$ production channel\dotfill}
& \multicolumn{4}{l}{VBF$_{1\gamma}$ production channel\dotfill} \\
& \multicolumn{2}{l}{HR$_{4b}$ reco.\ channel\dotfill}
& \multicolumn{2}{l}{HR$_{3b}$ reco.\ channel\dotfill} 
& \multicolumn{2}{l}{HR$_{4b}$ reco.\ channel\dotfill}
& \multicolumn{2}{l}{HR$_{3b}$ reco.\ channel\dotfill} \\
\sgline
& $S$ & $B$
& $S$ & $B$
& $S$ & $B$
& $S$ & $B$ \\
Total events for $150\,\textrm{fb}^{-1}$
& $5.7\times 10^5$   & \multicolumn{3}{l}{$6.5\times 10^{11}$\dotfill} 
& $1.4\times 10^4$   & \multicolumn{3}{l}{$2.9\times 10^8$\dotfill} \\
For $m_a=50\,\textrm{GeV}$ \\
\quad $n_{j}{=}6$ for HR$_{4b}$, $n_{j}{=}5$ for HR$_{3b}$
& $6.6\times 10^4$   & $3.5\times 10^9$
& $1.4\times 10^5$   & $9.2\times 10^9$
& $2096$             & $6.4\times 10^6$   & $3986$            & $1.4\times 10^7$ \\
\quad $n_{b}{=}4$ for HR$_{4b}$, $n_{b}{=}3$ for HR$_{3b}$  
& $965$              & $3.5\times 10^6$   & $1.2\times 10^4$  & $7.6\times 10^7$ 
& $29.8$             & $8661$             & $324$             & $1.5\times 10^5$ \\
\quad VBF production selections  
& $509$              & $4.8\times 10^5$   & $5146$            & $6.0\times 10^6$ 
& $8.6$              & $460$              & $91.2$            & $5028$ \\
\quad Higgs reconstruction selections 
& $295$              & $2.0\times 10^4$   & $2293$            & $2.5\times 10^5$
& $5.2$              & $9.9$              & $41.6$            & $178$ \\
For $m_a=35\,\textrm{GeV}$ \\
\quad $n_{j}{=}6$ for HR$_{4b}$, $n_{j}{=}5$ for HR$_{3b}$
& $6.3 \times 10^4$  & $3.5 \times 10^9$  & $1.4 \times 10^5$ & $9.2 \times 10^9$
& $1975$             & $6.4 \times 10^6$  & $3826$            & $1.4 \times 10^7$ \\
\quad $n_{b}{=}4$ for HR$_{4b}$, $n_{b}{=}3$ for HR$_{3b}$ 
& $994$              & $3.5 \times 10^6$  & $11934$           & $7.6 \times 10^7$
& $29$               & $8661$             & $309$             & $1.5 \times 10^5$ \\
\quad VBF production selections            
& $501$              & $4.8 \times 10^5$  & $4379$            & $6.0 \times 10^6$
& $8.2$              & $460$              & $77$              & $5028$ \\           
\quad Higgs reconstruction selections           
& $280$              & $7799$             & $1686$            & $2.2 \times 10^5$
& $4.6$              & $5.4$              & $30$              & $157$ \\
For $m_a=25\,\textrm{GeV}$ \\
\quad $n_{j}{=}6$ for HR$_{4b}$, $n_{j}{=}5$ for HR$_{3b}$
& $5.2 \times 10^4$  & $3.5 \times 10^9$  & $1.3 \times 10^5$ & $9.2 \times 10^9$
& $1427$             & $6.4 \times 10^6$  & $3155$            & $1.44 \times 10^7$ \\
\quad $n_{b}{=}4$ for HR$_{4b}$, $n_{b}{=}3$ for HR$_{3b}$ 
& $527$              & $3.5 \times 10^6$  & $8535$            & $7.6 \times 10^7$
& $13.5$             & $8661$             & $194$             & $1.5 \times 10^5$ \\
\quad VBF production selections            
& $165$              & $4.8 \times 10^5$  & $2455$            & $6.0 \times 10^6$
& $2.4$              & $460$              & $39$              & $5028$ \\           
\quad Higgs reconstruction selections           
& $72$               & $3518$             & $1522$            & $2.1 \times 10^5$
& $1.1$              & $0.8$              & $25$              & $150$ \\
For $m_a=10\,\textrm{GeV}$ \\
\quad $n_{j}{=}6$ for HR$_{4b}$, $n_{j}{=}5$ for HR$_{3b}$
& $3.6 \times 10^4$  & $3.5 \times 10^9$  & $1.1 \times 10^6$ & $9.2 \times 10^9$
& $1173$             & $6.4 \times 10^6$  & $3281$            & $1.44 \times 10^7$ \\
\quad $n_{b}{=}4$ for HR$_{4b}$, $n_{b}{=}3$ for HR$_{3b}$
& $42.7$              & $3.5 \times 10^6$  & $1186$            & $7.6 \times 10^7$
& $1.15$             & $8661$             & $41.4$             & $1.5 \times 10^5$ \\
\quad VBF production selections            
& $21.9$              & $4.8 \times 10^5$  & $419$            & $6.0 \times 10^6$
& $0.32$              & $460$              & $7.3$              & $5028$ \\           
\quad Higgs reconstruction selections           
& $7.44$               & $1609$             & $250$            & $3.2 \times 10^6$
& $0.02$              & $1.3$              & $4.01$              & $365$ \\
\dbline
\end{tabular}
}
\subtable[Signal sensitivity $S/\sqrt{B}$]{
\begin{tabular}{
p{0.280\textwidth}
p{0.080\textwidth}
p{0.080\textwidth}
p{0.080\textwidth}
p{0.080\textwidth}
p{0.080\textwidth}
p{0.080\textwidth}
p{0.080\textwidth}
p{0.080\textwidth}
}
\dbline
& \multicolumn{4}{l}{VBF$_{0\gamma}$ production channel\dotfill}
& \multicolumn{4}{l}{VBF$_{1\gamma}$ production channel\dotfill} \\
& \multicolumn{2}{l}{HR$_{4b}$ reco.\ channel\dotfill}
& \multicolumn{2}{l}{HR$_{3b}$ reco.\ channel\dotfill} 
& \multicolumn{2}{l}{HR$_{4b}$ reco.\ channel\dotfill}
& \multicolumn{2}{l}{HR$_{3b}$ reco.\ channel\dotfill} \\
\sgline
For $m_a=50\,\textrm{GeV}$ \\
\quad Per Higgs reconstruction channel
 &\multicolumn{2}{l}{$2.1\sigma$\dotfill}  
 &\multicolumn{2}{l}{$4.6\sigma$\dotfill}  
 &\multicolumn{2}{l}{$1.7\sigma$\dotfill}  
 &\multicolumn{2}{l}{$3.1\sigma$\dotfill}\\
\quad Per VBF production channel
 &\multicolumn{4}{l}{$5.1\sigma$\dotfill}  
 &\multicolumn{4}{l}{$3.5\sigma$\dotfill}\\
\quad All combined
 &\multicolumn{8}{l}{$6.1\sigma$\dotfill}\\
For $m_a=35\,\textrm{GeV}$ \\
\quad Per Higgs reconstruction channel
&\multicolumn{2}{l}{$3.2\sigma$\dotfill}  
&\multicolumn{2}{l}{$3.6\sigma$\dotfill}  
&\multicolumn{2}{l}{$2.0\sigma$\dotfill}  
&\multicolumn{2}{l}{$2.4\sigma$\dotfill}\\
\quad Per VBF production channel
&\multicolumn{4}{l}{$4.8\sigma$\dotfill}  
&\multicolumn{4}{l}{$3.1\sigma$\dotfill}\\
\quad All combined
&\multicolumn{8}{l}{$5.7\sigma$\dotfill}\\
For $m_a=25\,\textrm{GeV}$ \\
\quad Per Higgs reconstruction channel
 &\multicolumn{2}{l}{$1.2\sigma$\dotfill}  
 &\multicolumn{2}{l}{$3.3\sigma$\dotfill}  
 &\multicolumn{2}{l}{$1.2\sigma$\dotfill}  
 &\multicolumn{2}{l}{$2.0\sigma$\dotfill}\\
\quad Per VBF production channel
 &\multicolumn{4}{l}{$3.5\sigma$\dotfill}  
 &\multicolumn{4}{l}{$2.3\sigma$\dotfill}\\
\quad All combined
 &\multicolumn{8}{l}{$4.2\sigma$\dotfill}\\
For $m_a=10\,\textrm{GeV}$ \\
\quad Per Higgs reconstruction channel
&\multicolumn{2}{l}{$0.2\sigma$\dotfill}  
&\multicolumn{2}{l}{$0.4\sigma$\dotfill}  
&\multicolumn{2}{l}{$0.0\sigma$\dotfill}  
&\multicolumn{2}{l}{$0.2\sigma$\dotfill}\\
\quad Per VBF production channel
&\multicolumn{4}{l}{$0.5\sigma$\dotfill}  
&\multicolumn{4}{l}{$0.2\sigma$\dotfill}\\
\quad All combined
&\multicolumn{8}{l}{$0.5\sigma$\dotfill}\\
\dbline
\end{tabular}
}
}
\end{table*}

As noted in Sec.~\ref{sec:simulation}, samples are produced without pileup interactions. To investigate the effects of pileup on the signal sensitivity, a small sample of VBF$_{0\gamma}$ signal with a mean pileup of $\langle\mu\rangle=50$ is generated. For the VBF tagger, we find that $38\%$ of the pileup events pass compared to $39\%$ of the events without pileup. Similarly for the $O^{50}_{4b}$ Higgs tagger, we find that $50\%$ of the pileup events pass compared to $69\%$ of the events without pileup. The latter indicates that pileup may have up to a $30\%$ relative impact on the upper limit presented above.

\subsection{Trigger Optimization}
\label{sec:trigger}

We first compare various VBF jet trigger possibilities that affects both the VBF$_{0\gamma}$ and VBF$_{1\gamma}$ channels. Then we discuss the effect of the photon $p_\textrm{T}$ threshold for the VBF$_{1\gamma}$ channel.

\subsubsection*{VBF trigger comparisons}

In Sec.~\ref{sec:analysis}, we noted that optimistic assumptions were made about what data would be available from the trigger. Here, we revisit those assumptions, considering a range of alternate VBF triggers to compare to our BDT-based VBF tagger:
\begin{itemize}
\itemsep=0pt
\item ``Cuts,'' Our cut-based proposal made below on the VBF jets inspired by similar studies \cite{Sirunyan:2018kst,Aad:2020jym,Aaboud:2018gay,ATLAS:2020vbr,Carlson:2021tes}
\item ``ATLAS,'' ATLAS-inspired VBF trigger selections \cite{ATLAS:2021tnq,ATL-DAQ-PUB-2019-001}
\item ``CMS,'' CMS-inspired VBF trigger selections \cite{CMS:2020cmk}
\end{itemize}

As a comparison to the VBF BDT, we demonstrate cut-based selections on the VBF jets that could be implemented in the ATLAS trigger without using a BDT in the L1 trigger. 

Our cut-based proposal is as follows. 
\begin{itemize}
\itemsep=0pt
    \item $p_{\textrm{T},J1} > 50\,\textrm{GeV}$
    \item $p_{\textrm{T},J2} > 50\,\textrm{GeV}$
    \item $m_{JJ} > 1000\,\textrm{GeV}$ 
    \item $|\Delta \eta_{JJ}| > 3$
    \item $|\Delta \phi_{JJ}| < 2$
\end{itemize}

The ATLAS-inspired VBF trigger follows ATLAS's implementation that only utilizes the VBF jets, with no additional requirements, in the L1 subsystem that computes topological variables such as the invariant mass~\cite{ATLAS:2021tnq,ATL-DAQ-PUB-2019-001}. The set of offline selections, applied after the trigger, above which the trigger is greater than 95\% efficient is given by
\begin{itemize}
    \itemsep=0pt
    \item $p_{\textrm{T},J1} > 90\,\textrm{GeV}$
    \item $p_{\textrm{T},J2} > 80\,\textrm{GeV}$
    \item $|\eta_{J1}| < 3.2$
    \item $m_{JJ} > 1300\,\textrm{GeV}$
    \item $|\Delta \eta_{JJ}| > 4$
    \item $|\Delta \phi_{JJ}| < 2$
\end{itemize}
Approximately $40\,\textrm{fb}^{-1}$ of data with this trigger was collected during 2018~\cite{ATL-DAQ-PUB-2019-001}. We utilize the offline selections for our ATLAS-inspired L1 selections.

The CMS-inspired VBF trigger follows CMS's implementation of a similar VBF trigger \cite{CMS:2020cmk}. The set of offline trigger selections is given by
\begin{itemize}
    \itemsep=0pt
    \item $p_{\textrm{T},J1} > 150\,\textrm{GeV}$
    \item $p_{\textrm{T},J2} > 60\,\textrm{GeV}$
    \item $m_{JJ} > 800\,\textrm{GeV}$
\end{itemize}

A comparison of the results using the BDT-based trigger as well as the three introduced above is given in Table~\ref{tab:compare}. The table shows the sensitivity $\sigma$ using $S/\sqrt{B}$. The VBF BDT achieves $6\sigma$ while the others range from $2\sigma$ to $4\sigma$. This results confirms and expands on the previous findings of Ref.~\cite{Hong:2021snb}.

\begin{table*}[htb]
\caption{
    Sensitivity estimates for the $m_a=50\,\textrm{GeV}$ model assuming $\mathcal{B}(H_{4b}$) = 1, using $150\,\textrm{fb}^{-1}$ of data, the statistical sensitivity $S/\sqrt{B}$ of all final states after applying VBF and final state selections. Sensitivity is presented for using the ATLAS-inspired VBF HLT cut-based trigger, the CMS-inspired VBF L1 cut-based trigger, and our VBF cut-based trigger selections are presented here. Selections on the Higgs taggers corresponding to the HR channels are applied. For the VBF$_{1\gamma}$ channel, a photon $p_\textrm{T} \, > \, 15\, \textrm{GeV}$ threshold is used for all analyses.
    \label{tab:compare}
}
\centering
\begin{tabular}{
  p{0.320\textwidth}
  p{0.150\textwidth}
  p{0.150\textwidth}
  p{0.010\textwidth}
  p{0.150\textwidth}
  p{0.150\textwidth}
}
\dbline
&\multicolumn{2}{l}{VBF$_{0\gamma}$ production channel}
& 
&\multicolumn{2}{l}{VBF$_{1\gamma}$ production channel}\\
\cline{2-3}
\cline{5-6}
\clineskip
& HR$_{4b}$ reco.\ chan.
& HR$_{3b}$ reco.\ chan.
&
& HR$_{4b}$ reco.\ chan.
& HR$_{3b}$ reco.\ chan.
\\ 
\cline{1-3}
\cline{5-6}
\clineskip
Our VBF BDT (from Table \ref{tab:sigma}) \\
\qquad Per Higgs reconstruction channel
        & $2.1\sigma$
        & $4.6\sigma$
        &
        & $1.7\sigma$
        & $3.1\sigma$ \\
\qquad Per VBF production channel      
        & \multicolumn{2}{l}{$5.1\sigma$\dotfill}
        &
        & \multicolumn{2}{l}{$3.5\sigma$\dotfill} \\
\qquad Combined                        
        & \multicolumn{5}{l}{$6.1\sigma$\dotfill} \\
Our VBF cut-based trigger \\
\qquad Per Higgs reconstruction channel
        & $1.5\sigma$
        & $3.2\sigma$
        &
        & $1.5\sigma$
        & $2.5\sigma$ \\
\qquad Per VBF production channel      
        & \multicolumn{2}{l}{$3.5\sigma$\dotfill}
        &
        & \multicolumn{2}{l}{$2.9\sigma$\dotfill} \\
\qquad Combined                        
        & \multicolumn{5}{l}{$4.5\sigma$\dotfill} \\
ATLAS-inspired VBF trigger
\\
\qquad Per Higgs reconstruction channel
        & $0.8\sigma$
        & $1.7\sigma$
        &
        & $0.7\sigma$
        & $1.0\sigma$ \\
\qquad Per VBF production channel      
        & \multicolumn{2}{l}{$1.9\sigma$\dotfill}
        &
        & \multicolumn{2}{l}{$1.2\sigma$\dotfill} \\
\qquad Combined                        
        & \multicolumn{5}{l}{$2.2\sigma$\dotfill} \\
CMS-inspired VBF trigger \\
\qquad Per Higgs reconstruction channel
        & $0.9\sigma$
        & $1.6\sigma$
        &
        & $1.0 \sigma$
        & $1.0\sigma$ \\
\qquad Per VBF production channel      
        & \multicolumn{2}{l}{$1.8\sigma$\dotfill}
        &
        & \multicolumn{2}{l}{$1.4\sigma$\dotfill} \\
\qquad Combined                        
        & \multicolumn{5}{l}{$2.3\sigma$\dotfill} \\
\dbline
\end{tabular}

\end{table*}

\subsubsection*{Photon threshold for VBF trigger}

VBF$_{1\gamma}$ is a useful target for triggering because requiring a photon enables a significant reduction of the QCD multijet background. Therefore, with a photon, it is possible to implement a trigger with relatively low $p_\textrm{T}$ thresholds for VBF jets. As the current L1 trigger saves events with an electron or photon $p_\textrm{T} > $ 25 GeV~\cite{Aad:2019wsl}, many of these events are already saved by the trigger. Stricter selections can be used to further isolate signal at the HLT using VBF jets or $b$-tagging information.

For these reasons, VBF$_{1\gamma}$ has already been used in published analyses. For example, the ATLAS experiment searched for $H\rightarrow b\bar{b}$ with triggers requiring a photon with $p_\textrm{T} > 30\,\textrm{GeV}$, four or more jets with $p_\text{T} > 40\,\textrm{GeV}$, a VBF pair with $m_{JJ} > 700\,\textrm{GeV}$, and one or more $b$-tagged jet using 77\% efficiency selections~\cite{Aaboud:2018gay,ATLAS:2020vbr}. The trigger used in the study is similar to the selections described in Sec.~\ref{sec:vbf_tagger}; it obtains a similar sensitivity to that reported in this study. For the $H_{4b}$ search, however, a modification is necessary because the $p_\textrm{T} > 40\,\textrm{GeV}$ jet threshold significantly reduces signal acceptance. A modified approach could apply a tighter $m_{JJ}$ threshold with a lower jet $p_\textrm{T}$ threshold.  Additional details of the ATLAS trigger menu can be found in Refs.~\cite{ATL-DAQ-PUB-2016-001,ATL-DAQ-PUB-2017-001,ATL-DAQ-PUB-2018-002,ATL-DAQ-PUB-2019-001}.

The results shown so far in Tables~\ref{tab:sigma} and \ref{tab:compare} assume a threshold of photon $p_\textrm{T}>15\,\textrm{GeV}$.

\begin{figure}[htpb!]
\centering
\includegraphics[width=0.9\columnwidth]{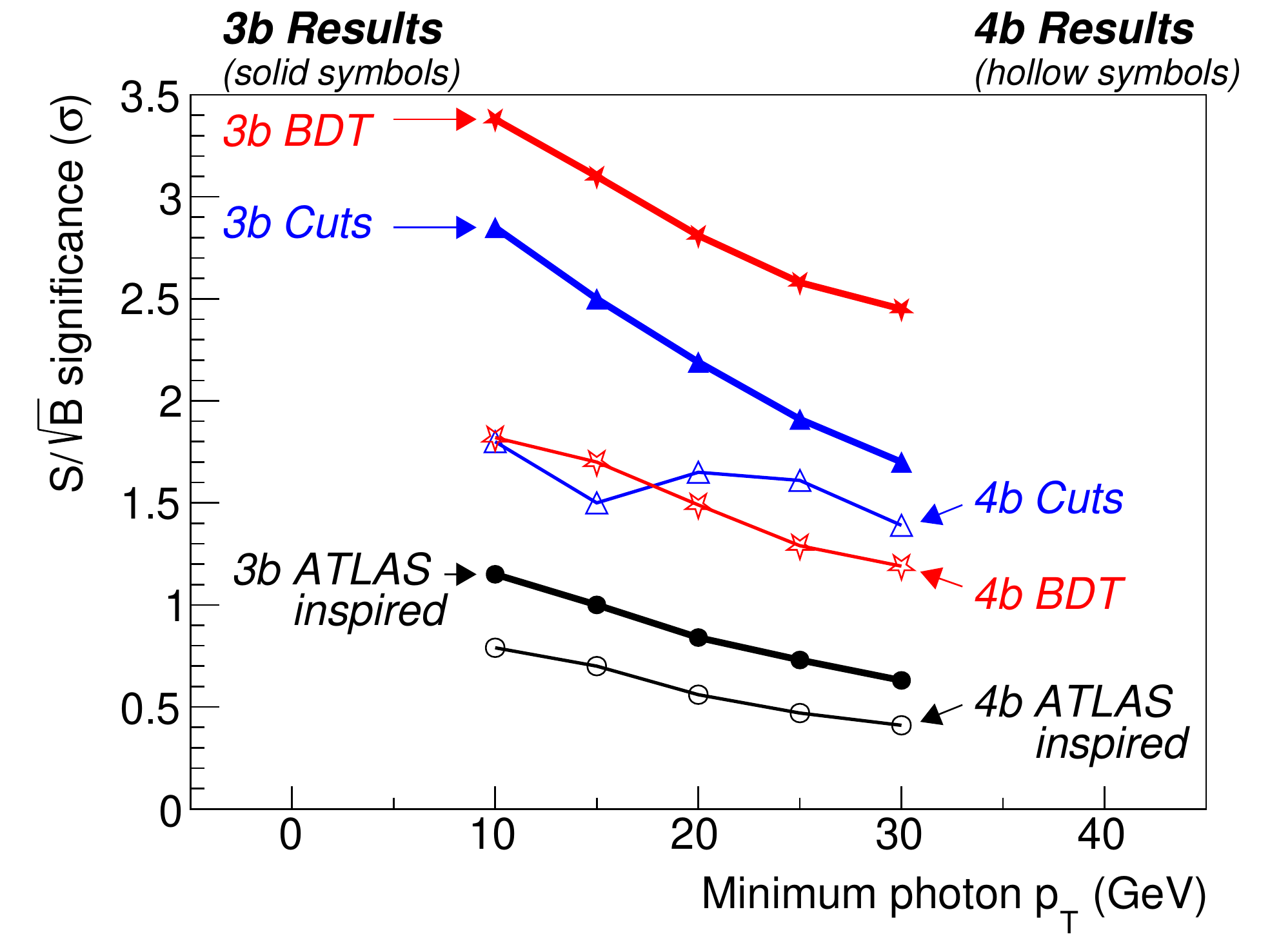}
\caption{
     Sensitivity vs.\ photon $p_\textrm{T}$ for the VBF$_{1\gamma}$ channel. The selections labeled ``ATLAS inspired'' refer to the selections using the ATLAS VBF trigger; see Section~\ref{sec:trigger}. A photon $p_\textrm{T} \, > \, 15\, \textrm{GeV}$ threshold is used for the nominal VBF$_{1\gamma}$ analysis in this paper.
    \label{fig:ypt}
}
\end{figure}

To evaluate the effect of varying the threshold of photon $p_\textrm{T}$, we evaluate the sensitivity as a function of photon $p_\textrm{T}$ for three VBF trigger scenarios: our VBF BDT trigger, our VBF cut-based trigger, and the ATLAS-inspired VBF trigger. The sensitivity as a function of photon $p_\textrm{T}$ is shown in Fig.~\ref{fig:ypt}. The plot shows that lower photon $p_\textrm{T}$ thresholds can increase the sensitivity by a factor of $1.5$ for the HR$_{3b}$ category. We leave the details and implementation to the experiments, but the studies here can be used as benchmarks.

\subsection{HL-LHC projection}
\label{sec:hl-lhc}

The HL-LHC is projected to collect approximately $\int \mathcal{L}\,dt = 3\,\textrm{ab}^{-1}$ of data, $20$ times the $150\,\textrm{fb}^{-1}$ assumed so far in this paper~\cite{Apollinari:2015wtw}. This increase in data improves the statistical sensitivity, resulting in an upper limit on branching ratio on the Higgs decay to four bottom quarks of $9\%$ in the VBF$_{0\gamma}$ channel and $13\%$ in the VBF$_{1\gamma}$ channel at the $95\%$ confidence level. A combination of the two production channels results in an upper limit on the branching ratio of $7\%$ at $95\%$ confidence level for $m_a$ = 50 GeV. There is an expected increase in $\sqrt{s}$ from $13.6\,\textrm{TeV}$ at the LHC Run-3 to $14\,\textrm{TeV}$ at HL-LHC. As stated previously for the increase from $13$ to $13.6\,\textrm{TeV}$, we do not expect the $\sqrt{s}$ increase to alter the conclusions presented here by more than a few relative percent.

\section{Conclusions}
\label{sec:conclusions}

We have presented a detailed study of VBF Higgs production with and without a photon in the $H\rightarrow aa\rightarrow 4b$ decay chain. To estimate the signal sensitivity, two relatively simple BDTs are deployed: one for the VBF jets for a ``VBF tagger,'' and one for the reconstructed Higgs decay for a ``Higgs tagger.'' Comparable sensitivity is found for the two VBF channels, with a combined sensitivity of over $6\sigma$ assuming a $150\,\textrm{fb}^{-1}$ dataset for $m_{a}=50\,\textrm{GeV}$.

The photon provides an additional way to retrieve events from the L1 trigger. We find that including the production channel with a photon yields a significantly improves combined sensitivity. We recommend that the LHC experiments provide combined triggers using photons and VBF jets in order to support this analysis for Run 3 and the HL-LHC.
The sensitivity of the proposed search was estimated using thresholds that may be implementable in the L1 trigger systems of ATLAS or CMS. To motivate these choices, we compare to existing triggers that are available in public documentation and provide benchmark selections to indicate the possible improvements. These results assume prompt decays of the $a$, but the conclusions regarding the trigger proposals are also relevant for 
longer lifetimes, such as those discussed in Ref.~\cite{Curtin:2015fna}.

A threshold-based ``cuts'' trigger yields sensitivity of up to $2.3\sigma$ for the current experiment triggers considered and up to $4.5\sigma$ for our proposed cut-based trigger. While these triggers produce a good sensitivity, the BDT-based trigger yields the best results. The availability of ML-on-FPGA packages allows for the implementation of BDT-based triggers at L1.
These tools will continue to enable improved sensitivity in searches for rare signals, such as the one presented here.

Looking forward, the development of low-latency and resource efficient ML implementations are paving way to more sophisticated constructions of artificial intelligence (AI) methods on FPGA, such as autoencoder-based anomaly detectors constructed using hls4ml-based neural networks \cite{Govorkova:2021utb} and \textsc{fwXmachina}-based decision trees \cite{Roche:2023int}. Such AI algorithms that were previously only executed in an offline environment may present opportunities for further improvements in the L1 trigger systems.

\section*{Acknowledgements}
We thank Gracie Jane Gollinger for computing infrastructure support at the University of Pittsburgh.
STR was supported by the Emil Sanielevici Undergraduate Research Scholarship.
BTC was supported by the National Science Foundation [award no.\ NSF-2209370].
CRH was supported by the US Department of Energy [award no.\ DE-SC0007859]. 
TMH was supported by the US Department of Energy [award no.\ DE-SC0007914].

\FloatBarrier
\normalem
\bibliography{bibliography.bib}

\begin{thebibliography}{117}%
\makeatletter
\providecommand \@ifxundefined [1]{%
 \@ifx{#1\undefined}
}%
\providecommand \@ifnum [1]{%
 \ifnum #1\expandafter \@firstoftwo
 \else \expandafter \@secondoftwo
 \fi
}%
\providecommand \@ifx [1]{%
 \ifx #1\expandafter \@firstoftwo
 \else \expandafter \@secondoftwo
 \fi
}%
\providecommand \natexlab [1]{#1}%
\providecommand \enquote  [1]{``#1''}%
\providecommand \bibnamefont  [1]{#1}%
\providecommand \bibfnamefont [1]{#1}%
\providecommand \citenamefont [1]{#1}%
\providecommand \href@noop [0]{\@secondoftwo}%
\providecommand \href [0]{\begingroup \@sanitize@url \@href}%
\providecommand \@href[1]{\@@startlink{#1}\@@href}%
\providecommand \@@href[1]{\endgroup#1\@@endlink}%
\providecommand \@sanitize@url [0]{\catcode `\\12\catcode `\$12\catcode
  `\&12\catcode `\#12\catcode `\^12\catcode `\_12\catcode `\%12\relax}%
\providecommand \@@startlink[1]{}%
\providecommand \@@endlink[0]{}%
\providecommand \url  [0]{\begingroup\@sanitize@url \@url }%
\providecommand \@url [1]{\endgroup\@href {#1}{\urlprefix }}%
\providecommand \urlprefix  [0]{URL }%
\providecommand \Eprint [0]{\href }%
\providecommand \doibase [0]{https://doi.org/}%
\providecommand \selectlanguage [0]{\@gobble}%
\providecommand \bibinfo  [0]{\@secondoftwo}%
\providecommand \bibfield  [0]{\@secondoftwo}%
\providecommand \translation [1]{[#1]}%
\providecommand \BibitemOpen [0]{}%
\providecommand \bibitemStop [0]{}%
\providecommand \bibitemNoStop [0]{.\EOS\space}%
\providecommand \EOS [0]{\spacefactor3000\relax}%
\providecommand \BibitemShut  [1]{\csname bibitem#1\endcsname}%
\let\auto@bib@innerbib\@empty
\bibitem [{\citenamefont {{ATLAS
  Collaboration}}(2012{\natexlab{a}})}]{Aad:2012tfa}%
  \BibitemOpen
  \bibfield  {author} {\bibinfo {author} {\bibnamefont {{ATLAS
  Collaboration}}},\ }\bibfield  {title} {\bibinfo {title} {{Observation of a
  new particle in the search for the Standard Model Higgs boson with the ATLAS
  detector at the LHC}},\ }\href
  {https://doi.org/10.1016/j.physletb.2012.08.020} {\bibfield  {journal}
  {\bibinfo  {journal} {Phys. Lett. B}\ }\textbf {\bibinfo {volume} {716}},\
  \bibinfo {pages} {1} (\bibinfo {year} {2012}{\natexlab{a}})}\BibitemShut
  {NoStop}%
\bibitem [{\citenamefont {{CMS Collaboration}}(2012)}]{Chatrchyan:2012ufa}%
  \BibitemOpen
  \bibfield  {author} {\bibinfo {author} {\bibnamefont {{CMS Collaboration}}},\
  }\bibfield  {title} {\bibinfo {title} {{Observation of a New Boson at a Mass
  of 125 GeV with the CMS Experiment at the LHC}},\ }\href
  {https://doi.org/10.1016/j.physletb.2012.08.021} {\bibfield  {journal}
  {\bibinfo  {journal} {Phys. Lett. B}\ }\textbf {\bibinfo {volume} {716}},\
  \bibinfo {pages} {30} (\bibinfo {year} {2012})}\BibitemShut {NoStop}%
\bibitem [{\citenamefont {{CMS
  Collaboration}}(2018{\natexlab{a}})}]{Sirunyan:2018kst}%
  \BibitemOpen
  \bibfield  {author} {\bibinfo {author} {\bibnamefont {{CMS Collaboration}}},\
  }\bibfield  {title} {\bibinfo {title} {{Observation of Higgs boson decay to
  bottom quarks}},\ }\href {https://doi.org/10.1103/PhysRevLett.121.121801}
  {\bibfield  {journal} {\bibinfo  {journal} {Phys. Rev. Lett.}\ }\textbf
  {\bibinfo {volume} {121}},\ \bibinfo {pages} {121801} (\bibinfo {year}
  {2018}{\natexlab{a}})}\BibitemShut {NoStop}%
\bibitem [{\citenamefont {{ATLAS
  Collaboration}}(2021{\natexlab{a}})}]{Aad:2020jym}%
  \BibitemOpen
  \bibfield  {author} {\bibinfo {author} {\bibnamefont {{ATLAS
  Collaboration}}},\ }\bibfield  {title} {\bibinfo {title} {{Measurements of
  $WH$ and $ZH$ production in the $H \rightarrow b\bar{b}$ decay channel in
  $pp$ collisions at 13 TeV with the ATLAS detector}},\ }\href
  {https://doi.org/10.1140/epjc/s10052-020-08677-2} {\bibfield  {journal}
  {\bibinfo  {journal} {Eur. Phys. J. C}\ }\textbf {\bibinfo {volume} {81}},\
  \bibinfo {pages} {178} (\bibinfo {year} {2021}{\natexlab{a}})}\BibitemShut
  {NoStop}%
\bibitem [{\citenamefont {{ATLAS
  Collaboration}}(2018{\natexlab{a}})}]{Aaboud:2018urx}%
  \BibitemOpen
  \bibfield  {author} {\bibinfo {author} {\bibnamefont {{ATLAS
  Collaboration}}},\ }\bibfield  {title} {\bibinfo {title} {{Observation of
  Higgs boson production in association with a top quark pair at the LHC with
  the ATLAS detector}},\ }\href
  {https://doi.org/10.1016/j.physletb.2018.07.035} {\bibfield  {journal}
  {\bibinfo  {journal} {Phys. Lett. B}\ }\textbf {\bibinfo {volume} {784}},\
  \bibinfo {pages} {173} (\bibinfo {year} {2018}{\natexlab{a}})}\BibitemShut
  {NoStop}%
\bibitem [{\citenamefont {{CMS
  Collaboration}}(2018{\natexlab{b}})}]{Sirunyan:2018hoz}%
  \BibitemOpen
  \bibfield  {author} {\bibinfo {author} {\bibnamefont {{CMS Collaboration}}},\
  }\bibfield  {title} {\bibinfo {title} {{Observation of
  $\mathrm{t\overline{t}}$H production}},\ }\href
  {https://doi.org/10.1103/PhysRevLett.120.231801} {\bibfield  {journal}
  {\bibinfo  {journal} {Phys. Rev. Lett.}\ }\textbf {\bibinfo {volume} {120}},\
  \bibinfo {pages} {231801} (\bibinfo {year} {2018}{\natexlab{b}})}\BibitemShut
  {NoStop}%
\bibitem [{\citenamefont {{ATLAS and CMS
  Collaborations}}(2016)}]{Khachatryan:2016vau}%
  \BibitemOpen
  \bibfield  {author} {\bibinfo {author} {\bibnamefont {{ATLAS and CMS
  Collaborations}}},\ }\bibfield  {title} {\bibinfo {title} {{Measurements of
  the Higgs boson production and decay rates and constraints on its couplings
  from a combined ATLAS and CMS analysis of the LHC \(pp\) collision data at
  \(\sqrt{s} = 7\) and \(8\,\text{TeV}\)}},\ }\href
  {https://doi.org/10.1007/JHEP08(2016)045} {\bibfield  {journal} {\bibinfo
  {journal} {J. High Energy Phys.}\ }\textbf {\bibinfo {volume} {08}}\bibinfo
  {number} { (2016)},\ \bibinfo {pages} {045}}\BibitemShut {NoStop}%
\bibitem [{\citenamefont {{ATLAS
  Collaboration}}(2019{\natexlab{a}})}]{Aaboud:2018pen}%
  \BibitemOpen
\bibfield  {number} {  }\bibfield  {author} {\bibinfo {author} {\bibnamefont
  {{ATLAS Collaboration}}},\ }\bibfield  {title} {\bibinfo {title}
  {{Cross-section measurements of the Higgs boson decaying into a pair of
  $\tau$-leptons in proton-proton collisions at $\sqrt{s}=13$ TeV with the
  ATLAS detector}},\ }\href {https://doi.org/10.1103/PhysRevD.99.072001}
  {\bibfield  {journal} {\bibinfo  {journal} {Phys. Rev. D}\ }\textbf {\bibinfo
  {volume} {99}},\ \bibinfo {pages} {072001} (\bibinfo {year}
  {2019}{\natexlab{a}})}\BibitemShut {NoStop}%
\bibitem [{\citenamefont {{CMS
  Collaboration}}(2018{\natexlab{c}})}]{Sirunyan:2017khh}%
  \BibitemOpen
  \bibfield  {author} {\bibinfo {author} {\bibnamefont {{CMS Collaboration}}},\
  }\bibfield  {title} {\bibinfo {title} {{Observation of the Higgs boson decay
  to a pair of $\tau$ leptons with the CMS detector}},\ }\href
  {https://doi.org/10.1016/j.physletb.2018.02.004} {\bibfield  {journal}
  {\bibinfo  {journal} {Phys. Lett. B}\ }\textbf {\bibinfo {volume} {779}},\
  \bibinfo {pages} {283} (\bibinfo {year} {2018}{\natexlab{c}})}\BibitemShut
  {NoStop}%
\bibitem [{\citenamefont {{ATLAS
  Collaboration}}(2021{\natexlab{b}})}]{ATLAS:2020fzp}%
  \BibitemOpen
  \bibfield  {author} {\bibinfo {author} {\bibnamefont {{ATLAS
  Collaboration}}},\ }\bibfield  {title} {\bibinfo {title} {{A search for the
  dimuon decay of the Standard Model Higgs boson with the ATLAS detector}},\
  }\href {https://doi.org/10.1016/j.physletb.2020.135980} {\bibfield  {journal}
  {\bibinfo  {journal} {Phys. Lett. B}\ }\textbf {\bibinfo {volume} {812}},\
  \bibinfo {pages} {135980} (\bibinfo {year} {2021}{\natexlab{b}})}\BibitemShut
  {NoStop}%
\bibitem [{\citenamefont {{ATLAS
  Collaboration}}(2022{\natexlab{a}})}]{ATLAS:2022ooq}%
  \BibitemOpen
  \bibfield  {author} {\bibinfo {author} {\bibnamefont {{ATLAS
  Collaboration}}},\ }\href@noop {} {\bibinfo {title} {{Measurements of Higgs
  boson production by gluon$-$gluon fusion and vector-boson fusion using
  $H\rightarrow W W^* \rightarrow e\nu \mu\nu$ decays in $pp$ collisions at
  $\sqrt{s}=13$ TeV with the ATLAS detector}}} (\bibinfo {year}
  {2022}{\natexlab{a}}),\ \Eprint {https://arxiv.org/abs/2207.00338}
  {arXiv:2207.00338 [hep-ex]} \BibitemShut {NoStop}%
\bibitem [{\citenamefont {{ATLAS
  Collaboration}}(2022{\natexlab{b}})}]{ATLAS:2021pkb}%
  \BibitemOpen
  \bibfield  {author} {\bibinfo {author} {\bibnamefont {{ATLAS
  Collaboration}}},\ }\bibfield  {title} {\bibinfo {title} {{Constraints on
  Higgs boson properties using $WW^{*}(\rightarrow e\nu \mu \nu )jj$ production
  in $36.1\,\mathrm{fb}^{-1}$ of $\sqrt{s}=13$~TeV pp collisions with the ATLAS
  detector}},\ }\href {https://doi.org/10.1140/epjc/s10052-022-10366-1}
  {\bibfield  {journal} {\bibinfo  {journal} {Eur. Phys. J. C}\ }\textbf
  {\bibinfo {volume} {82}},\ \bibinfo {pages} {622} (\bibinfo {year}
  {2022}{\natexlab{b}})}\BibitemShut {NoStop}%
\bibitem [{\citenamefont {{ATLAS
  Collaboration}}(2020{\natexlab{a}})}]{ATLAS:2020ior}%
  \BibitemOpen
  \bibfield  {author} {\bibinfo {author} {\bibnamefont {{ATLAS
  Collaboration}}},\ }\bibfield  {title} {\bibinfo {title} {{$CP$ Properties of
  Higgs Boson Interactions with Top Quarks in the $t\bar{t}H$ and $tH$
  Processes Using $H \rightarrow \gamma\gamma$ with the ATLAS Detector}},\
  }\href {https://doi.org/10.1103/PhysRevLett.125.061802} {\bibfield  {journal}
  {\bibinfo  {journal} {Phys. Rev. Lett.}\ }\textbf {\bibinfo {volume} {125}},\
  \bibinfo {pages} {061802} (\bibinfo {year} {2020}{\natexlab{a}})}\BibitemShut
  {NoStop}%
\bibitem [{\citenamefont {{ATLAS
  Collaboration}}(2020{\natexlab{b}})}]{ATLAS:2020evk}%
  \BibitemOpen
  \bibfield  {author} {\bibinfo {author} {\bibnamefont {{ATLAS
  Collaboration}}},\ }\bibfield  {title} {\bibinfo {title} {{Test of CP
  invariance in vector-boson fusion production of the Higgs boson in the $H
  \rightarrow\tau \tau$ channel in proton-proton collisions at s=13 TeV with
  the ATLAS detector}},\ }\href
  {https://doi.org/10.1016/j.physletb.2020.135426} {\bibfield  {journal}
  {\bibinfo  {journal} {Phys. Lett. B}\ }\textbf {\bibinfo {volume} {805}},\
  \bibinfo {pages} {135426} (\bibinfo {year} {2020}{\natexlab{b}})}\BibitemShut
  {NoStop}%
\bibitem [{\citenamefont {{ATLAS
  Collaboration}}(2022{\natexlab{c}})}]{ATLAS:2022tan}%
  \BibitemOpen
  \bibfield  {author} {\bibinfo {author} {\bibnamefont {{ATLAS
  Collaboration}}},\ }\href@noop {} {\bibinfo {title} {{Study of the CP
  property of the Higgs boson to electroweak boson coupling in the VBF
  $H\to\gamma\gamma$ channel with the ATLAS detector}}} (\bibinfo {year}
  {2022}{\natexlab{c}}),\ \Eprint {https://arxiv.org/abs/2208.02338}
  {arXiv:2208.02338 [hep-ex]} \BibitemShut {NoStop}%
\bibitem [{\citenamefont {{CMS
  Collaboration}}(2022{\natexlab{a}})}]{CMS:2021sdq}%
  \BibitemOpen
  \bibfield  {author} {\bibinfo {author} {\bibnamefont {{CMS Collaboration}}},\
  }\bibfield  {title} {\bibinfo {title} {{Analysis of the $CP$ structure of the
  Yukawa coupling between the Higgs boson and $\tau$ leptons in proton-proton
  collisions at $ \sqrt{s} $ = 13 TeV}},\ }\href
  {https://doi.org/10.1007/JHEP06(2022)012} {\bibfield  {journal} {\bibinfo
  {journal} {J. High Energy Phys.}\ }\textbf {\bibinfo {volume} {06}}\bibinfo
  {number} { (2022)},\ \bibinfo {pages} {012}}\BibitemShut {NoStop}%
\bibitem [{\citenamefont {{CMS
  Collaboration}}(2022{\natexlab{b}})}]{CMS:2022dbt}%
  \BibitemOpen
\bibfield  {number} {  }\bibfield  {author} {\bibinfo {author} {\bibnamefont
  {{CMS Collaboration}}},\ }\href@noop {} {\bibinfo {title} {{Search for $CP$
  violation in ttH and tH production in multilepton channels in proton-proton
  collisions at $\sqrt{s}$ = 13 TeV}}} (\bibinfo {year} {2022}{\natexlab{b}}),\
  \Eprint {https://arxiv.org/abs/2208.02686} {arXiv:2208.02686 [hep-ex]}
  \BibitemShut {NoStop}%
\bibitem [{\citenamefont {{ATLAS
  Collaboration}}(2022{\natexlab{d}})}]{ATLAS:2022vkf}%
  \BibitemOpen
  \bibfield  {author} {\bibinfo {author} {\bibnamefont {{ATLAS
  Collaboration}}},\ }\bibfield  {title} {\bibinfo {title} {{A detailed map of
  Higgs boson interactions by the ATLAS experiment ten years after the
  discovery}},\ }\href {https://doi.org/10.1038/s41586-022-04893-w} {\bibfield
  {journal} {\bibinfo  {journal} {Nature}\ }\textbf {\bibinfo {volume} {607}},\
  \bibinfo {pages} {52} (\bibinfo {year} {2022}{\natexlab{d}})},\ \bibinfo
  {note} {[Erratum: Nature 612, E24 (2022)]}\BibitemShut {NoStop}%
\bibitem [{\citenamefont {{CMS
  Collaboration}}(2022{\natexlab{c}})}]{CMS:2022dwd}%
  \BibitemOpen
  \bibfield  {author} {\bibinfo {author} {\bibnamefont {{CMS Collaboration}}},\
  }\bibfield  {title} {\bibinfo {title} {{A portrait of the Higgs boson by the
  CMS experiment ten years after the discovery}},\ }\href
  {https://doi.org/10.1038/s41586-022-04892-x} {\bibfield  {journal} {\bibinfo
  {journal} {Nature}\ }\textbf {\bibinfo {volume} {607}},\ \bibinfo {pages}
  {60} (\bibinfo {year} {2022}{\natexlab{c}})}\BibitemShut {NoStop}%
\bibitem [{\citenamefont {Curtin}\ \emph {et~al.}(2014)\citenamefont {Curtin}
  \emph {et~al.}}]{Curtin:2013fra}%
  \BibitemOpen
  \bibfield  {author} {\bibinfo {author} {\bibfnamefont {D.}~\bibnamefont
  {Curtin}} \emph {et~al.},\ }\bibfield  {title} {\bibinfo {title} {{Exotic
  decays of the 125 GeV Higgs boson}},\ }\href
  {https://doi.org/10.1103/PhysRevD.90.075004} {\bibfield  {journal} {\bibinfo
  {journal} {Phys. Rev. D}\ }\textbf {\bibinfo {volume} {90}},\ \bibinfo
  {pages} {075004} (\bibinfo {year} {2014})}\BibitemShut {NoStop}%
\bibitem [{\citenamefont {Shrock}\ and\ \citenamefont
  {Suzuki}(1982)}]{SHROCK1982250}%
  \BibitemOpen
  \bibfield  {author} {\bibinfo {author} {\bibfnamefont {R.~E.}\ \bibnamefont
  {Shrock}}\ and\ \bibinfo {author} {\bibfnamefont {M.}~\bibnamefont
  {Suzuki}},\ }\bibfield  {title} {\bibinfo {title} {{Invisible decays of Higgs
  bosons}},\ }\href
  {https://doi.org/https://doi.org/10.1016/0370-2693(82)91247-3} {\bibfield
  {journal} {\bibinfo  {journal} {Phys. Lett. B}\ }\textbf {\bibinfo {volume}
  {110}},\ \bibinfo {pages} {250} (\bibinfo {year} {1982})}\BibitemShut
  {NoStop}%
\bibitem [{\citenamefont {Dobrescu}\ and\ \citenamefont
  {Matchev}(2000)}]{Dobrescu:2000yn}%
  \BibitemOpen
  \bibfield  {author} {\bibinfo {author} {\bibfnamefont {B.~A.}\ \bibnamefont
  {Dobrescu}}\ and\ \bibinfo {author} {\bibfnamefont {K.~T.}\ \bibnamefont
  {Matchev}},\ }\bibfield  {title} {\bibinfo {title} {{Light axion within the
  next-to-minimal supersymmetric standard model}},\ }\href
  {https://doi.org/10.1088/1126-6708/2000/09/031} {\bibfield  {journal}
  {\bibinfo  {journal} {J. High Energy Phys.}\ }\textbf {\bibinfo {volume}
  {09}}\bibinfo  {number} { (2000)},\ \bibinfo {pages} {031}}\BibitemShut
  {NoStop}%
\bibitem [{\citenamefont {Ellwanger}\ \emph {et~al.}(2003)\citenamefont
  {Ellwanger}, \citenamefont {Gunion}, \citenamefont {Hugonie},\ and\
  \citenamefont {Moretti}}]{Ellwanger:2003jt}%
  \BibitemOpen
\bibfield  {number} {  }\bibfield  {author} {\bibinfo {author} {\bibfnamefont
  {U.}~\bibnamefont {Ellwanger}}, \bibinfo {author} {\bibfnamefont {J.~F.}\
  \bibnamefont {Gunion}}, \bibinfo {author} {\bibfnamefont {C.}~\bibnamefont
  {Hugonie}},\ and\ \bibinfo {author} {\bibfnamefont {S.}~\bibnamefont
  {Moretti}},\ }\href@noop {} {\bibinfo {title} {{Towards a no lose theorem for
  NMSSM Higgs discovery at the LHC}}} (\bibinfo {year} {2003}),\ \Eprint
  {https://arxiv.org/abs/hep-ph/0305109} {arXiv:hep-ph/0305109} \BibitemShut
  {NoStop}%
\bibitem [{\citenamefont {Ellwanger}\ \emph
  {et~al.}(2005{\natexlab{a}})\citenamefont {Ellwanger}, \citenamefont
  {Gunion},\ and\ \citenamefont {Hugonie}}]{Ellwanger:2005uu}%
  \BibitemOpen
  \bibfield  {author} {\bibinfo {author} {\bibfnamefont {U.}~\bibnamefont
  {Ellwanger}}, \bibinfo {author} {\bibfnamefont {J.~F.}\ \bibnamefont
  {Gunion}},\ and\ \bibinfo {author} {\bibfnamefont {C.}~\bibnamefont
  {Hugonie}},\ }\bibfield  {title} {\bibinfo {title} {{Difficult scenarios for
  NMSSM Higgs discovery at the LHC}},\ }\href
  {https://doi.org/10.1088/1126-6708/2005/07/041} {\bibfield  {journal}
  {\bibinfo  {journal} {J. High Energy Phys.}\ }\textbf {\bibinfo {volume}
  {07}}\bibinfo  {number} { (2005)},\ \bibinfo {pages} {041}}\BibitemShut
  {NoStop}%
\bibitem [{\citenamefont {Stelzer}\ \emph {et~al.}(2007)\citenamefont
  {Stelzer}, \citenamefont {Wiesenfeldt},\ and\ \citenamefont
  {Willenbrock}}]{Stelzer:2006sp}%
  \BibitemOpen
\bibfield  {number} {  }\bibfield  {author} {\bibinfo {author} {\bibfnamefont
  {T.}~\bibnamefont {Stelzer}}, \bibinfo {author} {\bibfnamefont
  {S.}~\bibnamefont {Wiesenfeldt}},\ and\ \bibinfo {author} {\bibfnamefont
  {S.}~\bibnamefont {Willenbrock}},\ }\bibfield  {title} {\bibinfo {title}
  {{Higgs at the Tevatron in Extended Supersymmetric Models}},\ }\href
  {https://doi.org/10.1103/PhysRevD.75.077701} {\bibfield  {journal} {\bibinfo
  {journal} {Phys. Rev. D}\ }\textbf {\bibinfo {volume} {75}},\ \bibinfo
  {pages} {077701} (\bibinfo {year} {2007})}\BibitemShut {NoStop}%
\bibitem [{\citenamefont {Cheung}\ \emph {et~al.}(2007)\citenamefont {Cheung},
  \citenamefont {Song},\ and\ \citenamefont {Yan}}]{Cheung:2007sva}%
  \BibitemOpen
  \bibfield  {author} {\bibinfo {author} {\bibfnamefont {K.}~\bibnamefont
  {Cheung}}, \bibinfo {author} {\bibfnamefont {J.}~\bibnamefont {Song}},\ and\
  \bibinfo {author} {\bibfnamefont {Q.-S.}\ \bibnamefont {Yan}},\ }\bibfield
  {title} {\bibinfo {title} {Role of
  $h\ensuremath{\rightarrow}\ensuremath{\eta}\ensuremath{\eta}$ in
  intermediate-mass higgs boson searches at the large hadron collider},\ }\href
  {https://doi.org/10.1103/PhysRevLett.99.031801} {\bibfield  {journal}
  {\bibinfo  {journal} {Phys. Rev. Lett.}\ }\textbf {\bibinfo {volume} {99}},\
  \bibinfo {pages} {031801} (\bibinfo {year} {2007})}\BibitemShut {NoStop}%
\bibitem [{\citenamefont {Carena}\ \emph {et~al.}(2008)\citenamefont {Carena},
  \citenamefont {Han}, \citenamefont {Huang},\ and\ \citenamefont
  {Wagner}}]{Carena:2007jk}%
  \BibitemOpen
  \bibfield  {author} {\bibinfo {author} {\bibfnamefont {M.}~\bibnamefont
  {Carena}}, \bibinfo {author} {\bibfnamefont {T.}~\bibnamefont {Han}},
  \bibinfo {author} {\bibfnamefont {G.-Y.}\ \bibnamefont {Huang}},\ and\
  \bibinfo {author} {\bibfnamefont {C.~E.}\ \bibnamefont {Wagner}},\ }\bibfield
   {title} {\bibinfo {title} {{Higgs Signal for h $\to$ aa at Hadron
  Colliders}},\ }\href {https://doi.org/10.1088/1126-6708/2008/04/092}
  {\bibfield  {journal} {\bibinfo  {journal} {J. High Energy Phys.}\ }\textbf
  {\bibinfo {volume} {04}}\bibinfo  {number} { (2008)},\ \bibinfo {pages}
  {092}}\BibitemShut {NoStop}%
\bibitem [{\citenamefont {Chang}\ \emph {et~al.}(2008)\citenamefont {Chang},
  \citenamefont {Dermisek}, \citenamefont {Gunion},\ and\ \citenamefont
  {Weiner}}]{Chang:2008cw}%
  \BibitemOpen
\bibfield  {number} {  }\bibfield  {author} {\bibinfo {author} {\bibfnamefont
  {S.}~\bibnamefont {Chang}}, \bibinfo {author} {\bibfnamefont
  {R.}~\bibnamefont {Dermisek}}, \bibinfo {author} {\bibfnamefont {J.~F.}\
  \bibnamefont {Gunion}},\ and\ \bibinfo {author} {\bibfnamefont
  {N.}~\bibnamefont {Weiner}},\ }\bibfield  {title} {\bibinfo {title}
  {{Nonstandard Higgs Boson Decays}},\ }\href
  {https://doi.org/10.1146/annurev.nucl.58.110707.171200} {\bibfield  {journal}
  {\bibinfo  {journal} {Ann. Rev. Nucl. Part. Sci.}\ }\textbf {\bibinfo
  {volume} {58}},\ \bibinfo {pages} {75} (\bibinfo {year} {2008})}\BibitemShut
  {NoStop}%
\bibitem [{\citenamefont {Cao}\ \emph {et~al.}(2013)\citenamefont {Cao},
  \citenamefont {Ding}, \citenamefont {Han}, \citenamefont {Yang},\ and\
  \citenamefont {Zhu}}]{Cao:2013gba}%
  \BibitemOpen
  \bibfield  {author} {\bibinfo {author} {\bibfnamefont {J.}~\bibnamefont
  {Cao}}, \bibinfo {author} {\bibfnamefont {F.}~\bibnamefont {Ding}}, \bibinfo
  {author} {\bibfnamefont {C.}~\bibnamefont {Han}}, \bibinfo {author}
  {\bibfnamefont {J.~M.}\ \bibnamefont {Yang}},\ and\ \bibinfo {author}
  {\bibfnamefont {J.}~\bibnamefont {Zhu}},\ }\bibfield  {title} {\bibinfo
  {title} {{A light Higgs scalar in the NMSSM confronted with the latest LHC
  Higgs data}},\ }\href {https://doi.org/10.1007/JHEP11(2013)018} {\bibfield
  {journal} {\bibinfo  {journal} {J. High Energy Phys.}\ }\textbf {\bibinfo
  {volume} {11}}\bibinfo  {number} { (2013)},\ \bibinfo {pages}
  {018}}\BibitemShut {NoStop}%
\bibitem [{\citenamefont {Cepeda}\ \emph {et~al.}(2021)\citenamefont {Cepeda},
  \citenamefont {Gori}, \citenamefont {Outschoorn},\ and\ \citenamefont
  {Shelton}}]{Cepeda:2021rql}%
  \BibitemOpen
\bibfield  {number} {  }\bibfield  {author} {\bibinfo {author} {\bibfnamefont
  {M.}~\bibnamefont {Cepeda}}, \bibinfo {author} {\bibfnamefont
  {S.}~\bibnamefont {Gori}}, \bibinfo {author} {\bibfnamefont {V.~M.}\
  \bibnamefont {Outschoorn}},\ and\ \bibinfo {author} {\bibfnamefont
  {J.}~\bibnamefont {Shelton}},\ }\bibfield  {title} {\bibinfo {title} {{Exotic
  Higgs Decays}},\ }\href
  {https://www.annualreviews.org/doi/10.1146/annurev-nucl-102319-024147}
  {\bibfield  {journal} {\bibinfo  {journal} {Ann. Rev. of Nucl. and Part.
  Sci.}\ }\textbf {\bibinfo {volume} {72}},\ \bibinfo {pages} {119} (\bibinfo
  {year} {2021})}\BibitemShut {NoStop}%
\bibitem [{\citenamefont {Kaplan}\ and\ \citenamefont
  {McEvoy}(2011{\natexlab{a}})}]{Kaplan:2011vf}%
  \BibitemOpen
  \bibfield  {author} {\bibinfo {author} {\bibfnamefont {D.~E.}\ \bibnamefont
  {Kaplan}}\ and\ \bibinfo {author} {\bibfnamefont {M.}~\bibnamefont
  {McEvoy}},\ }\bibfield  {title} {\bibinfo {title} {{Associated Production of
  Non-Standard Higgs Bosons at the LHC}},\ }\href
  {https://doi.org/10.1103/PhysRevD.83.115004} {\bibfield  {journal} {\bibinfo
  {journal} {Phys. Rev. D}\ }\textbf {\bibinfo {volume} {83}},\ \bibinfo
  {pages} {115004} (\bibinfo {year} {2011}{\natexlab{a}})}\BibitemShut
  {NoStop}%
\bibitem [{\citenamefont {Kaplan}\ and\ \citenamefont
  {McEvoy}(2011{\natexlab{b}})}]{Kaplan:2009qt}%
  \BibitemOpen
  \bibfield  {author} {\bibinfo {author} {\bibfnamefont {D.~E.}\ \bibnamefont
  {Kaplan}}\ and\ \bibinfo {author} {\bibfnamefont {M.}~\bibnamefont
  {McEvoy}},\ }\bibfield  {title} {\bibinfo {title} {{Searching for Higgs
  decays to four bottom quarks at LHCb}},\ }\href
  {https://doi.org/10.1016/j.physletb.2011.05.026} {\bibfield  {journal}
  {\bibinfo  {journal} {Phys. Lett. B}\ }\textbf {\bibinfo {volume} {701}},\
  \bibinfo {pages} {70} (\bibinfo {year} {2011}{\natexlab{b}})}\BibitemShut
  {NoStop}%
\bibitem [{\citenamefont {Bomark}\ \emph {et~al.}(2016)\citenamefont {Bomark},
  \citenamefont {Moretti},\ and\ \citenamefont {Roszkowski}}]{Bomark:2015fga}%
  \BibitemOpen
  \bibfield  {author} {\bibinfo {author} {\bibfnamefont {N.-E.}\ \bibnamefont
  {Bomark}}, \bibinfo {author} {\bibfnamefont {S.}~\bibnamefont {Moretti}},\
  and\ \bibinfo {author} {\bibfnamefont {L.}~\bibnamefont {Roszkowski}},\
  }\bibfield  {title} {\bibinfo {title} {{Detection prospects of light NMSSM
  Higgs pseudoscalar via cascades of heavier scalars from vector boson fusion
  and Higgs-strahlung}},\ }\href
  {https://doi.org/10.1088/0954-3899/43/10/105003} {\bibfield  {journal}
  {\bibinfo  {journal} {J. Phys. G}\ }\textbf {\bibinfo {volume} {43}},\
  \bibinfo {pages} {105003} (\bibinfo {year} {2016})}\BibitemShut {NoStop}%
\bibitem [{\citenamefont {Ellwanger}\ \emph
  {et~al.}(2005{\natexlab{b}})\citenamefont {Ellwanger}, \citenamefont
  {Gunion},\ and\ \citenamefont {Hugonie}}]{Ellwanger_2005}%
  \BibitemOpen
  \bibfield  {author} {\bibinfo {author} {\bibfnamefont {U.}~\bibnamefont
  {Ellwanger}}, \bibinfo {author} {\bibfnamefont {J.~F.}\ \bibnamefont
  {Gunion}},\ and\ \bibinfo {author} {\bibfnamefont {C.}~\bibnamefont
  {Hugonie}},\ }\bibfield  {title} {\bibinfo {title} {Difficult scenarios for
  {NMSSM} higgs discovery at the {LHC}},\ }\href
  {https://doi.org/10.1088/1126-6708/2005/07/041} {\bibfield  {journal}
  {\bibinfo  {journal} {J. High Energy Phys.}\ }\textbf {\bibinfo {volume}
  {07}}\bibinfo  {number} { (2005)},\ \bibinfo {pages} {041}}\BibitemShut
  {NoStop}%
\bibitem [{\citenamefont {Dawson}\ \emph {et~al.}(2008)\citenamefont {Dawson}
  \emph {et~al.}}]{dawson2008higgs}%
  \BibitemOpen
\bibfield  {number} {  }\bibfield  {author} {\bibinfo {author} {\bibfnamefont
  {S.}~\bibnamefont {Dawson}} \emph {et~al.},\ }\href@noop {} {\bibinfo {title}
  {{Higgs Working Group Summary Report}}} (\bibinfo {year} {2008}),\ \Eprint
  {https://arxiv.org/abs/0803.1154} {arXiv:0803.1154 [hep-ph]} \BibitemShut
  {NoStop}%
\bibitem [{\citenamefont {Curtin}\ and\ \citenamefont
  {Verhaaren}(2015)}]{Curtin:2015fna}%
  \BibitemOpen
  \bibfield  {author} {\bibinfo {author} {\bibfnamefont {D.}~\bibnamefont
  {Curtin}}\ and\ \bibinfo {author} {\bibfnamefont {C.~B.}\ \bibnamefont
  {Verhaaren}},\ }\bibfield  {title} {\bibinfo {title} {{Discovering Uncolored
  Naturalness in Exotic Higgs Decays}},\ }\href
  {https://doi.org/10.1007/JHEP12(2015)072} {\bibfield  {journal} {\bibinfo
  {journal} {JHEP}\ }\textbf {\bibinfo {volume} {12}}\bibinfo  {number} {
  (2015)},\ \bibinfo {pages} {072}}\BibitemShut {NoStop}%
\bibitem [{\citenamefont {Adhikary}\ \emph {et~al.}(2022)\citenamefont
  {Adhikary}, \citenamefont {Banerjee}, \citenamefont {Barman}, \citenamefont
  {Batell}, \citenamefont {Bhattacherjee}, \citenamefont {Bose}, \citenamefont
  {Qian},\ and\ \citenamefont {Spannowsky}}]{Adhikary:2022jfp}%
  \BibitemOpen
\bibfield  {number} {  }\bibfield  {author} {\bibinfo {author} {\bibfnamefont
  {A.}~\bibnamefont {Adhikary}}, \bibinfo {author} {\bibfnamefont
  {S.}~\bibnamefont {Banerjee}}, \bibinfo {author} {\bibfnamefont {R.~K.}\
  \bibnamefont {Barman}}, \bibinfo {author} {\bibfnamefont {B.}~\bibnamefont
  {Batell}}, \bibinfo {author} {\bibfnamefont {B.}~\bibnamefont
  {Bhattacherjee}}, \bibinfo {author} {\bibfnamefont {C.}~\bibnamefont {Bose}},
  \bibinfo {author} {\bibfnamefont {Z.}~\bibnamefont {Qian}},\ and\ \bibinfo
  {author} {\bibfnamefont {M.}~\bibnamefont {Spannowsky}},\ }\href@noop {}
  {\bibinfo {title} {{Prospects for Exotic $h\rightarrow 4 \tau$ Decays in
  Single and Di-Higgs Production at the LHC and Future Hadron Colliders}}}
  (\bibinfo {year} {2022}),\ \Eprint {https://arxiv.org/abs/2211.07674}
  {arXiv:2211.07674 [hep-ph]} \BibitemShut {NoStop}%
\bibitem [{\citenamefont {Martin}(2007)}]{Martin:2007dx}%
  \BibitemOpen
  \bibfield  {author} {\bibinfo {author} {\bibfnamefont {A.}~\bibnamefont
  {Martin}},\ }\href@noop {} {\bibinfo {title} {{Higgs Cascade Decays to gamma
  gamma + jet jet at the LHC}}} (\bibinfo {year} {2007}),\ \Eprint
  {https://arxiv.org/abs/hep-ph/0703247} {arXiv:hep-ph/0703247} \BibitemShut
  {NoStop}%
\bibitem [{\citenamefont {{ATLAS
  Collaboration}}(2018{\natexlab{b}})}]{ATLAS:2018jnf}%
  \BibitemOpen
  \bibfield  {author} {\bibinfo {author} {\bibnamefont {{ATLAS
  Collaboration}}},\ }\bibfield  {title} {\bibinfo {title} {{Search for Higgs
  boson decays into pairs of light (pseudo)scalar particles in the
  $\gamma\gamma jj$ final state in $pp$ collisions at $\sqrt{s}=13$ TeV with
  the ATLAS detector}},\ }\href
  {https://doi.org/10.1016/j.physletb.2018.06.011} {\bibfield  {journal}
  {\bibinfo  {journal} {Phys. Lett. B}\ }\textbf {\bibinfo {volume} {782}},\
  \bibinfo {pages} {750} (\bibinfo {year} {2018}{\natexlab{b}})}\BibitemShut
  {NoStop}%
\bibitem [{\citenamefont {{ATLAS
  Collaboration}}(2018{\natexlab{c}})}]{ATLAS:2018pvw}%
  \BibitemOpen
  \bibfield  {author} {\bibinfo {author} {\bibnamefont {{ATLAS
  Collaboration}}},\ }\bibfield  {title} {\bibinfo {title} {{Search for the
  Higgs boson produced in association with a vector boson and decaying into two
  spin-zero particles in the $H \rightarrow aa \rightarrow 4b$ channel in $pp$
  collisions at $\sqrt{s} = 13$ TeV with the ATLAS detector}},\ }\href
  {https://doi.org/10.1007/JHEP10(2018)031} {\bibfield  {journal} {\bibinfo
  {journal} {J. High Energy Phys.}\ }\textbf {\bibinfo {volume} {10}}\bibinfo
  {number} { (2018)},\ \bibinfo {pages} {031}}\BibitemShut {NoStop}%
\bibitem [{\citenamefont {{ATLAS
  Collaboration}}(2020{\natexlab{c}})}]{ATLAS:2020ahi}%
  \BibitemOpen
\bibfield  {number} {  }\bibfield  {author} {\bibinfo {author} {\bibnamefont
  {{ATLAS Collaboration}}},\ }\bibfield  {title} {\bibinfo {title} {{Search for
  Higgs boson decays into two new low-mass spin-0 particles in the 4$b$ channel
  with the ATLAS detector using $pp$ collisions at $\sqrt{s}= 13$ TeV}},\
  }\href {https://doi.org/10.1103/PhysRevD.102.112006} {\bibfield  {journal}
  {\bibinfo  {journal} {Phys. Rev. D}\ }\textbf {\bibinfo {volume} {102}},\
  \bibinfo {pages} {112006} (\bibinfo {year} {2020}{\natexlab{c}})}\BibitemShut
  {NoStop}%
\bibitem [{\citenamefont {{ATLAS
  Collaboration}}(2020{\natexlab{d}})}]{ATLAS:2020jgy}%
  \BibitemOpen
  \bibfield  {author} {\bibinfo {author} {\bibnamefont {{ATLAS
  Collaboration}}},\ }\bibfield  {title} {\bibinfo {title} {{Search for the $HH
  \rightarrow b \bar{b} b \bar{b}$ process via vector-boson fusion production
  using proton-proton collisions at $\sqrt{s} = 13$ TeV with the ATLAS
  detector}},\ }\href {https://doi.org/10.1007/JHEP07(2020)108} {\bibfield
  {journal} {\bibinfo  {journal} {J. High Energy Phys.}\ }\textbf {\bibinfo
  {volume} {07}}\bibfield  {number} {\bibinfo  {number} { (2020)},\ \bibinfo
  {pages} {108}},\ }\bibinfo {note} {[Erratum: J. High Energy Phys. \textbf{01}
  (2021), 145, Erratum: J. High Energy Phys. \textbf{05} (2021),
  207]}\BibitemShut {NoStop}%
\bibitem [{\citenamefont {{ATLAS
  Collaboration}}(2022{\natexlab{e}})}]{ATLAS:2021hbr}%
  \BibitemOpen
  \bibfield  {author} {\bibinfo {author} {\bibnamefont {{ATLAS
  Collaboration}}},\ }\bibfield  {title} {\bibinfo {title} {{Search for Higgs
  boson decays into a pair of pseudoscalar particles in the $bb\mu\mu$ final
  state with the ATLAS detector in $pp$ collisions at $\sqrt s$=13\,\,TeV}},\
  }\href {https://doi.org/10.1103/PhysRevD.105.012006} {\bibfield  {journal}
  {\bibinfo  {journal} {Phys. Rev. D}\ }\textbf {\bibinfo {volume} {105}},\
  \bibinfo {pages} {012006} (\bibinfo {year} {2022}{\natexlab{e}})}\BibitemShut
  {NoStop}%
\bibitem [{\citenamefont {{CMS Collaboration}}(2019)}]{CMS:2018nsh}%
  \BibitemOpen
  \bibfield  {author} {\bibinfo {author} {\bibnamefont {{CMS Collaboration}}},\
  }\bibfield  {title} {\bibinfo {title} {{Search for an exotic decay of the
  Higgs boson to a pair of light pseudoscalars in the final state with two
  muons and two b quarks in pp collisions at 13 TeV}},\ }\href
  {https://doi.org/10.1016/j.physletb.2019.06.021} {\bibfield  {journal}
  {\bibinfo  {journal} {Phys. Lett. B}\ }\textbf {\bibinfo {volume} {795}},\
  \bibinfo {pages} {398} (\bibinfo {year} {2019})}\BibitemShut {NoStop}%
\bibitem [{\citenamefont {{CMS
  Collaboration}}(2023{\natexlab{a}})}]{CMS-PAS-HIG-22-007}%
  \BibitemOpen
  \bibfield  {author} {\bibinfo {author} {\bibnamefont {{CMS Collaboration}}},\
  }\href@noop {} {\bibinfo {title} {{Search for exotic Higgs boson decays to a
  pair of pseudoscalars in the $\mu\mu\mathrm{bb}$ and $\tau\tau\mathrm{bb}$
  final states in proton-proton collisions with the CMS experiment}}},\
  \bibinfo {howpublished}
  {\href{https://cds.cern.ch/record/2853298}{CMS-PAS-HIG-22-007}} (\bibinfo
  {year} {2023}{\natexlab{a}})\BibitemShut {NoStop}%
\bibitem [{\citenamefont {Haisch}\ \emph {et~al.}(2018)\citenamefont {Haisch},
  \citenamefont {Kamenik}, \citenamefont {Malinauskas},\ and\ \citenamefont
  {Spira}}]{Haisch:2018kqx}%
  \BibitemOpen
  \bibfield  {author} {\bibinfo {author} {\bibfnamefont {U.}~\bibnamefont
  {Haisch}}, \bibinfo {author} {\bibfnamefont {J.~F.}\ \bibnamefont {Kamenik}},
  \bibinfo {author} {\bibfnamefont {A.}~\bibnamefont {Malinauskas}},\ and\
  \bibinfo {author} {\bibfnamefont {M.}~\bibnamefont {Spira}},\ }\bibfield
  {title} {\bibinfo {title} {{Collider constraints on light pseudoscalars}},\
  }\href {https://doi.org/10.1007/JHEP03(2018)178} {\bibfield  {journal}
  {\bibinfo  {journal} {J. High Energy Phys.}\ }\textbf {\bibinfo {volume}
  {03}}\bibinfo  {number} { (2018)},\ \bibinfo {pages} {178}}\BibitemShut
  {NoStop}%
\bibitem [{\citenamefont {Baer}\ \emph {et~al.}(2013)\citenamefont {Baer} \emph
  {et~al.}}]{Baer:2013cma}%
  \BibitemOpen
\bibfield  {number} {  }\bibfield  {author} {\bibinfo {author} {\bibfnamefont
  {H.}~\bibnamefont {Baer}} \emph {et~al.},\ }\href@noop {} {\bibinfo {title}
  {{The International Linear Collider Technical Design Report - Volume 2:
  Physics}}} (\bibinfo {year} {2013}),\ \Eprint
  {https://arxiv.org/abs/1306.6352} {arXiv:1306.6352 [hep-ph]} \BibitemShut
  {NoStop}%
\bibitem [{\citenamefont {Agostini}\ \emph {et~al.}(2021)\citenamefont
  {Agostini} \emph {et~al.}}]{LHeC:2020van}%
  \BibitemOpen
  \bibfield  {author} {\bibinfo {author} {\bibfnamefont {P.}~\bibnamefont
  {Agostini}} \emph {et~al.} (\bibinfo {collaboration} {LHeC, FCC-he Study
  Group}),\ }\bibfield  {title} {\bibinfo {title} {{The Large
  Hadron\textendash{}Electron Collider at the HL-LHC}},\ }\href
  {https://doi.org/10.1088/1361-6471/abf3ba} {\bibfield  {journal} {\bibinfo
  {journal} {J. Phys. G}\ }\textbf {\bibinfo {volume} {48}},\ \bibinfo {pages}
  {110501} (\bibinfo {year} {2021})}\BibitemShut {NoStop}%
\bibitem [{\citenamefont {Liu}\ \emph {et~al.}(2017{\natexlab{a}})\citenamefont
  {Liu}, \citenamefont {Tang}, \citenamefont {Zhang},\ and\ \citenamefont
  {Zhu}}]{Liu:2016ahc}%
  \BibitemOpen
  \bibfield  {author} {\bibinfo {author} {\bibfnamefont {S.}~\bibnamefont
  {Liu}}, \bibinfo {author} {\bibfnamefont {Y.-L.}\ \bibnamefont {Tang}},
  \bibinfo {author} {\bibfnamefont {C.}~\bibnamefont {Zhang}},\ and\ \bibinfo
  {author} {\bibfnamefont {S.-h.}\ \bibnamefont {Zhu}},\ }\bibfield  {title}
  {\bibinfo {title} {{Exotic Higgs Decay $h\rightarrow\phi\phi\rightarrow 4b$
  at the LHeC}},\ }\href {https://doi.org/10.1140/epjc/s10052-017-5012-5}
  {\bibfield  {journal} {\bibinfo  {journal} {Eur. Phys. J. C}\ }\textbf
  {\bibinfo {volume} {77}},\ \bibinfo {pages} {457} (\bibinfo {year}
  {2017}{\natexlab{a}})}\BibitemShut {NoStop}%
\bibitem [{\citenamefont {Liu}\ \emph {et~al.}(2017{\natexlab{b}})\citenamefont
  {Liu}, \citenamefont {Wang},\ and\ \citenamefont {Zhang}}]{Liu:2016zki}%
  \BibitemOpen
  \bibfield  {author} {\bibinfo {author} {\bibfnamefont {Z.}~\bibnamefont
  {Liu}}, \bibinfo {author} {\bibfnamefont {L.-T.}\ \bibnamefont {Wang}},\ and\
  \bibinfo {author} {\bibfnamefont {H.}~\bibnamefont {Zhang}},\ }\bibfield
  {title} {\bibinfo {title} {{Exotic decays of the 125 GeV Higgs boson at
  future $e^+e^-$ lepton colliders}},\ }\href
  {https://doi.org/10.1088/1674-1137/41/6/063102} {\bibfield  {journal}
  {\bibinfo  {journal} {Chin. Phys. C}\ }\textbf {\bibinfo {volume} {41}},\
  \bibinfo {pages} {063102} (\bibinfo {year} {2017}{\natexlab{b}})}\BibitemShut
  {NoStop}%
\bibitem [{\citenamefont {Carena}\ \emph {et~al.}(2023)\citenamefont {Carena},
  \citenamefont {Kozaczuk}, \citenamefont {Liu}, \citenamefont {Ou},
  \citenamefont {Ramsey-Musolf}, \citenamefont {Shelton}, \citenamefont
  {Wang},\ and\ \citenamefont {Xie}}]{Carena:2022yvx}%
  \BibitemOpen
  \bibfield  {author} {\bibinfo {author} {\bibfnamefont {M.}~\bibnamefont
  {Carena}}, \bibinfo {author} {\bibfnamefont {J.}~\bibnamefont {Kozaczuk}},
  \bibinfo {author} {\bibfnamefont {Z.}~\bibnamefont {Liu}}, \bibinfo {author}
  {\bibfnamefont {T.}~\bibnamefont {Ou}}, \bibinfo {author} {\bibfnamefont
  {M.~J.}\ \bibnamefont {Ramsey-Musolf}}, \bibinfo {author} {\bibfnamefont
  {J.}~\bibnamefont {Shelton}}, \bibinfo {author} {\bibfnamefont
  {Y.}~\bibnamefont {Wang}},\ and\ \bibinfo {author} {\bibfnamefont {K.-P.}\
  \bibnamefont {Xie}},\ }\bibfield  {title} {\bibinfo {title} {{Probing the
  Electroweak Phase Transition with Exotic Higgs Decays}},\ }\href
  {https://doi.org/10.31526/lhep.2023.432} {\bibfield  {journal} {\bibinfo
  {journal} {LHEP}\ }\textbf {\bibinfo {volume} {2023}},\ \bibinfo {pages}
  {432} (\bibinfo {year} {2023})}\BibitemShut {NoStop}%
\bibitem [{\citenamefont {Gabrielli}\ \emph {et~al.}(2007)\citenamefont
  {Gabrielli}, \citenamefont {Maltoni}, \citenamefont {Mele}, \citenamefont
  {Moretti}, \citenamefont {Piccinini},\ and\ \citenamefont
  {Pittau}}]{Gabrielli:2007wf}%
  \BibitemOpen
  \bibfield  {author} {\bibinfo {author} {\bibfnamefont {E.}~\bibnamefont
  {Gabrielli}}, \bibinfo {author} {\bibfnamefont {F.}~\bibnamefont {Maltoni}},
  \bibinfo {author} {\bibfnamefont {B.}~\bibnamefont {Mele}}, \bibinfo {author}
  {\bibfnamefont {M.}~\bibnamefont {Moretti}}, \bibinfo {author} {\bibfnamefont
  {F.}~\bibnamefont {Piccinini}},\ and\ \bibinfo {author} {\bibfnamefont
  {R.}~\bibnamefont {Pittau}},\ }\bibfield  {title} {\bibinfo {title} {{Higgs
  Boson Production in Association with a Photon in Vector Boson Fusion at the
  LHC}},\ }\href {https://doi.org/10.1016/j.nuclphysb.2007.05.010} {\bibfield
  {journal} {\bibinfo  {journal} {Nucl. Phys. B}\ }\textbf {\bibinfo {volume}
  {781}},\ \bibinfo {pages} {64} (\bibinfo {year} {2007})}\BibitemShut
  {NoStop}%
\bibitem [{\citenamefont {Gabrielli}\ \emph {et~al.}(2016)\citenamefont
  {Gabrielli}, \citenamefont {Mele}, \citenamefont {Piccinini},\ and\
  \citenamefont {Pittau}}]{Gabrielli:2016mdd}%
  \BibitemOpen
  \bibfield  {author} {\bibinfo {author} {\bibfnamefont {E.}~\bibnamefont
  {Gabrielli}}, \bibinfo {author} {\bibfnamefont {B.}~\bibnamefont {Mele}},
  \bibinfo {author} {\bibfnamefont {F.}~\bibnamefont {Piccinini}},\ and\
  \bibinfo {author} {\bibfnamefont {R.}~\bibnamefont {Pittau}},\ }\bibfield
  {title} {\bibinfo {title} {{Asking for an extra photon in Higgs production at
  the LHC and beyond}},\ }\href {https://doi.org/10.1007/JHEP07(2016)003}
  {\bibfield  {journal} {\bibinfo  {journal} {J. High Energy Phys.}\ }\textbf
  {\bibinfo {volume} {07}}\bibinfo  {number} { (2016)},\ \bibinfo {pages}
  {003}}\BibitemShut {NoStop}%
\bibitem [{\citenamefont {Asner}\ \emph {et~al.}(2010)\citenamefont {Asner},
  \citenamefont {Cunningham}, \citenamefont {Dejong}, \citenamefont
  {Randrianarivony}, \citenamefont {Rios},\ and\ \citenamefont
  {Schram}}]{Asner:2010mj}%
  \BibitemOpen
\bibfield  {number} {  }\bibfield  {author} {\bibinfo {author} {\bibfnamefont
  {D.}~\bibnamefont {Asner}}, \bibinfo {author} {\bibfnamefont
  {M.}~\bibnamefont {Cunningham}}, \bibinfo {author} {\bibfnamefont
  {S.}~\bibnamefont {Dejong}}, \bibinfo {author} {\bibfnamefont
  {K.}~\bibnamefont {Randrianarivony}}, \bibinfo {author} {\bibfnamefont
  {C.}~\bibnamefont {Rios}},\ and\ \bibinfo {author} {\bibfnamefont
  {M.}~\bibnamefont {Schram}},\ }\href@noop {} {\bibinfo {title} {{Search for a
  Light Standard Model Higgs Boson Produced in Association with a Photon in
  Vector Boson Fusion}}} (\bibinfo {year} {2010}),\ \Eprint
  {https://arxiv.org/abs/1006.0669} {arXiv:1006.0669 [hep-ph]} \BibitemShut
  {NoStop}%
\bibitem [{\citenamefont {Carlson}\ \emph {et~al.}(2021)\citenamefont
  {Carlson}, \citenamefont {Han},\ and\ \citenamefont
  {Leung}}]{Carlson:2021tes}%
  \BibitemOpen
  \bibfield  {author} {\bibinfo {author} {\bibfnamefont {B.}~\bibnamefont
  {Carlson}}, \bibinfo {author} {\bibfnamefont {T.}~\bibnamefont {Han}},\ and\
  \bibinfo {author} {\bibfnamefont {S.~C.~I.}\ \bibnamefont {Leung}},\
  }\bibfield  {title} {\bibinfo {title} {{Higgs boson to charm quark decay in
  vector boson fusion plus a photon}},\ }\href
  {https://doi.org/10.1103/PhysRevD.104.073006} {\bibfield  {journal} {\bibinfo
   {journal} {Phys. Rev. D}\ }\textbf {\bibinfo {volume} {104}},\ \bibinfo
  {pages} {073006} (\bibinfo {year} {2021})}\BibitemShut {NoStop}%
\bibitem [{\citenamefont {{ATLAS
  Collaboration}}(2018{\natexlab{d}})}]{Aaboud:2018gay}%
  \BibitemOpen
  \bibfield  {author} {\bibinfo {author} {\bibnamefont {{ATLAS
  Collaboration}}},\ }\bibfield  {title} {\bibinfo {title} {{Search for Higgs
  bosons produced via vector-boson fusion and decaying into bottom quark pairs
  in $\sqrt{s} = 13$ $\mathrm{TeV}$ $pp$ collisions with the ATLAS detector}},\
  }\href {https://doi.org/10.1103/PhysRevD.98.052003} {\bibfield  {journal}
  {\bibinfo  {journal} {Phys. Rev. D}\ }\textbf {\bibinfo {volume} {98}},\
  \bibinfo {pages} {052003} (\bibinfo {year} {2018}{\natexlab{d}})}\BibitemShut
  {NoStop}%
\bibitem [{\citenamefont {{ATLAS
  Collaboration}}(2021{\natexlab{c}})}]{ATLAS:2020cvh}%
  \BibitemOpen
  \bibfield  {author} {\bibinfo {author} {\bibnamefont {{ATLAS
  Collaboration}}},\ }\bibfield  {title} {\bibinfo {title} {{Search for Higgs
  boson production in association with a high-energy photon via vector-boson
  fusion with decay into bottom quark pairs at $\sqrt{s}$=13 TeV with the ATLAS
  detector}},\ }\href {https://doi.org/10.1007/JHEP03(2021)268} {\bibfield
  {journal} {\bibinfo  {journal} {J. High Energy Phys.}\ }\textbf {\bibinfo
  {volume} {03}}\bibinfo  {number} { (2021)},\ \bibinfo {pages}
  {268}}\BibitemShut {NoStop}%
\bibitem [{\citenamefont {{ATLAS
  Collaboration}}(2022{\natexlab{f}})}]{ATLAS:2021pdg}%
  \BibitemOpen
\bibfield  {number} {  }\bibfield  {author} {\bibinfo {author} {\bibnamefont
  {{ATLAS Collaboration}}},\ }\bibfield  {title} {\bibinfo {title}
  {{Observation of electroweak production of two jets in association with an
  isolated photon and missing transverse momentum, and search for a Higgs boson
  decaying into invisible particles at 13~$\text {TeV}$ with the ATLAS
  detector}},\ }\href {https://doi.org/10.1140/epjc/s10052-021-09878-z}
  {\bibfield  {journal} {\bibinfo  {journal} {Eur. Phys. J. C}\ }\textbf
  {\bibinfo {volume} {82}},\ \bibinfo {pages} {105} (\bibinfo {year}
  {2022}{\natexlab{f}})}\BibitemShut {NoStop}%
\bibitem [{\citenamefont {{ATLAS Collaboration}}(2023)}]{ATLAS:2023tkt}%
  \BibitemOpen
  \bibfield  {author} {\bibinfo {author} {\bibnamefont {{ATLAS
  Collaboration}}},\ }\bibfield  {title} {\bibinfo {title} {{Combination of
  searches for invisible decays of the Higgs boson using
  \(139\,\text{fb}^{-1}\) of proton--proton collision data at \(\sqrt{s} =
  13\,\text{TeV}\) collected with the ATLAS experiment}},\ }\href
  {https://doi.org/10.1016/j.physletb.2023.137963} {\bibfield  {journal}
  {\bibinfo  {journal} {Phys. Lett. B}\ }\textbf {\bibinfo {volume} {842}},\
  \bibinfo {pages} {137963} (\bibinfo {year} {2023})}\BibitemShut {NoStop}%
\bibitem [{\citenamefont {{CMS
  Collaboration}}(2023{\natexlab{b}})}]{CMS:2023sdw}%
  \BibitemOpen
  \bibfield  {author} {\bibinfo {author} {\bibnamefont {{CMS Collaboration}}}
  (\bibinfo {collaboration} {CMS}),\ }\bibfield  {title} {\bibinfo {title} {{A
  search for decays of the Higgs boson to invisible particles in events with a
  top-antitop quark pair or a vector boson in proton-proton collisions at
  $\sqrt{s} = 13\,\text {Te}\hspace{-.08em}\text {V} $}},\ }\href
  {https://doi.org/10.1140/epjc/s10052-023-11952-7} {\bibfield  {journal}
  {\bibinfo  {journal} {Eur. Phys. J. C}\ }\textbf {\bibinfo {volume} {83}},\
  \bibinfo {pages} {933} (\bibinfo {year} {2023}{\natexlab{b}})}\BibitemShut
  {NoStop}%
\bibitem [{\citenamefont {Duarte}\ \emph {et~al.}(2018)\citenamefont {Duarte}
  \emph {et~al.}}]{Duarte:2018ite}%
  \BibitemOpen
  \bibfield  {author} {\bibinfo {author} {\bibfnamefont {J.}~\bibnamefont
  {Duarte}} \emph {et~al.},\ }\bibfield  {title} {\bibinfo {title} {{Fast
  inference of deep neural networks in FPGAs for particle physics}},\ }\href
  {https://doi.org/10.1088/1748-0221/13/07/P07027} {\bibfield  {journal}
  {\bibinfo  {journal} {J. Instrum.}\ }\textbf {\bibinfo {volume} {13}}\bibinfo
   {number} { (07)},\ \bibinfo {pages} {P07027}}\BibitemShut {NoStop}%
\bibitem [{\citenamefont {Loncar}\ \emph {et~al.}(2021)\citenamefont {Loncar}
  \emph {et~al.}}]{Loncar:2020hqp}%
  \BibitemOpen
\bibfield  {number} {  }\bibfield  {author} {\bibinfo {author} {\bibfnamefont
  {V.}~\bibnamefont {Loncar}} \emph {et~al.},\ }\bibfield  {title} {\bibinfo
  {title} {{Compressing deep neural networks on FPGAs to binary and ternary
  precision with HLS4ML}},\ }\href {https://doi.org/10.1088/2632-2153/aba042}
  {\bibfield  {journal} {\bibinfo  {journal} {Mach. Learn. Sci. Tech.}\
  }\textbf {\bibinfo {volume} {2}},\ \bibinfo {pages} {015001} (\bibinfo {year}
  {2021})}\BibitemShut {NoStop}%
\bibitem [{\citenamefont {Aarrestad}\ \emph {et~al.}(2021)\citenamefont
  {Aarrestad} \emph {et~al.}}]{Aarrestad:2021zos}%
  \BibitemOpen
  \bibfield  {author} {\bibinfo {author} {\bibfnamefont {T.}~\bibnamefont
  {Aarrestad}} \emph {et~al.},\ }\bibfield  {title} {\bibinfo {title} {{Fast
  convolutional neural networks on FPGAs with hls4ml}},\ }\href
  {https://doi.org/10.1088/2632-2153/ac0ea1} {\bibfield  {journal} {\bibinfo
  {journal} {Mach. Learn. Sci. Tech.}\ }\textbf {\bibinfo {volume} {2}},\
  \bibinfo {pages} {045015} (\bibinfo {year} {2021})}\BibitemShut {NoStop}%
\bibitem [{\citenamefont {Khoda}\ \emph {et~al.}(2023)\citenamefont {Khoda}
  \emph {et~al.}}]{Khoda:2022dwz}%
  \BibitemOpen
  \bibfield  {author} {\bibinfo {author} {\bibfnamefont {E.~E.}\ \bibnamefont
  {Khoda}} \emph {et~al.},\ }\bibfield  {title} {\bibinfo {title} {{Ultra-low
  latency recurrent neural network inference on FPGAs for physics applications
  with hls4ml}},\ }\href {https://doi.org/10.1088/2632-2153/acc0d7} {\bibfield
  {journal} {\bibinfo  {journal} {Mach. Learn. Sci. Tech.}\ }\textbf {\bibinfo
  {volume} {4}},\ \bibinfo {pages} {025004} (\bibinfo {year}
  {2023})}\BibitemShut {NoStop}%
\bibitem [{\citenamefont {Summers}\ \emph {et~al.}(2020)\citenamefont {Summers}
  \emph {et~al.}}]{Summers:2020xiy}%
  \BibitemOpen
  \bibfield  {author} {\bibinfo {author} {\bibfnamefont {S.}~\bibnamefont
  {Summers}} \emph {et~al.},\ }\bibfield  {title} {\bibinfo {title} {{Fast
  inference of Boosted Decision Trees in FPGAs for particle physics}},\ }\href
  {https://doi.org/10.1088/1748-0221/15/05/P05026} {\bibfield  {journal}
  {\bibinfo  {journal} {J. Instrum.}\ }\textbf {\bibinfo {volume} {15}}\bibinfo
   {number} { (05)},\ \bibinfo {pages} {P05026}}\BibitemShut {NoStop}%
\bibitem [{\citenamefont {{T.M. Hong, B.T. Carlson, B. Eubanks, S. Racz, S.
  Roche, J. Stelzer, and D. Stumpp}}(2021)}]{Hong:2021snb}%
  \BibitemOpen
\bibfield  {number} {  }\bibfield  {author} {\bibinfo {author} {\bibnamefont
  {{T.M. Hong, B.T. Carlson, B. Eubanks, S. Racz, S. Roche, J. Stelzer, and D.
  Stumpp}}},\ }\bibfield  {title} {\bibinfo {title} {{Nanosecond machine
  learning event classification with boosted decision trees in FPGA for high
  energy physics}},\ }\href {https://doi.org/10.1088/1748-0221/16/08/P08016}
  {\bibfield  {journal} {\bibinfo  {journal} {J. Instrum.}\ }\textbf {\bibinfo
  {volume} {16}}\bibinfo  {number} { (08)},\ \bibinfo {pages}
  {P08016}}\BibitemShut {NoStop}%
\bibitem [{\citenamefont {Carlson}\ \emph {et~al.}(2022)\citenamefont
  {Carlson}, \citenamefont {Bayer}, \citenamefont {Hong},\ and\ \citenamefont
  {Roche}}]{Carlson:2022dgb}%
  \BibitemOpen
\bibfield  {number} {  }\bibfield  {author} {\bibinfo {author} {\bibfnamefont
  {B.}~\bibnamefont {Carlson}}, \bibinfo {author} {\bibfnamefont
  {Q.}~\bibnamefont {Bayer}}, \bibinfo {author} {\bibfnamefont {T.~M.}\
  \bibnamefont {Hong}},\ and\ \bibinfo {author} {\bibfnamefont
  {S.}~\bibnamefont {Roche}},\ }\bibfield  {title} {\bibinfo {title}
  {{Nanosecond machine learning regression with deep boosted decision trees in
  FPGA for high energy physics}},\ }\href
  {https://doi.org/10.1088/1748-0221/17/09/P09039} {\bibfield  {journal}
  {\bibinfo  {journal} {J. Instrum.}\ }\textbf {\bibinfo {volume} {17}}\bibinfo
   {number} { (09)},\ \bibinfo {pages} {P09039}}\BibitemShut {NoStop}%
\bibitem [{\citenamefont {Achenbach}\ \emph {et~al.}(2008)\citenamefont
  {Achenbach} \emph {et~al.}}]{Achenbach:2008zzb}%
  \BibitemOpen
\bibfield  {number} {  }\bibfield  {author} {\bibinfo {author} {\bibfnamefont
  {R.}~\bibnamefont {Achenbach}} \emph {et~al.},\ }\bibfield  {title} {\bibinfo
  {title} {{The ATLAS level-1 calorimeter trigger}},\ }\href
  {https://doi.org/10.1088/1748-0221/3/03/P03001} {\bibfield  {journal}
  {\bibinfo  {journal} {J. Instrum.}\ }\textbf {\bibinfo {volume} {3}},\
  \bibinfo {pages} {P03001}}\BibitemShut {NoStop}%
\bibitem [{\citenamefont {{CMS
  Collaboration}}(2020{\natexlab{a}})}]{CMS:2020cmk}%
  \BibitemOpen
  \bibfield  {author} {\bibinfo {author} {\bibnamefont {{CMS Collaboration}}},\
  }\bibfield  {title} {\bibinfo {title} {{Performance of the CMS Level-1
  trigger in proton-proton collisions at $\sqrt{s} =$ 13 TeV}},\ }\href
  {https://doi.org/10.1088/1748-0221/15/10/P10017} {\bibfield  {journal}
  {\bibinfo  {journal} {J. Instrum.}\ }\textbf {\bibinfo {volume} {15}}\bibinfo
   {number} { (10)},\ \bibinfo {pages} {P10017}}\BibitemShut {NoStop}%
\bibitem [{\citenamefont {{ATLAS Collaboration}}(2016)}]{ATL-DAQ-PUB-2016-001}%
  \BibitemOpen
\bibfield  {number} {  }\bibfield  {author} {\bibinfo {author} {\bibnamefont
  {{ATLAS Collaboration}}},\ }\href@noop {} {\bibinfo {title} {{2015 start-up
  trigger menu and initial performance assessment of the ATLAS trigger using
  Run-2 data}}},\ \bibinfo {howpublished}
  {\href{https://cds.cern.ch/record/2136007}{ATL-DAQ-PUB-2016-001}} (\bibinfo
  {year} {2016})\BibitemShut {NoStop}%
\bibitem [{\citenamefont {{ATLAS
  Collaboration}}(2017{\natexlab{a}})}]{ATL-DAQ-PUB-2017-001}%
  \BibitemOpen
  \bibfield  {author} {\bibinfo {author} {\bibnamefont {{ATLAS
  Collaboration}}},\ }\href@noop {} {\bibinfo {title} {{Trigger Menu in
  2016}}},\ \bibinfo {howpublished}
  {\href{https://cds.cern.ch/record/2242069}{ATL-DAQ-PUB-2017-001}} (\bibinfo
  {year} {2017}{\natexlab{a}})\BibitemShut {NoStop}%
\bibitem [{\citenamefont {{ATLAS
  Collaboration}}(2018{\natexlab{e}})}]{ATL-DAQ-PUB-2018-002}%
  \BibitemOpen
  \bibfield  {author} {\bibinfo {author} {\bibnamefont {{ATLAS
  Collaboration}}},\ }\href@noop {} {\bibinfo {title} {{Trigger Menu in
  2017}}},\ \bibinfo {howpublished}
  {\href{https://cds.cern.ch/record/2625986}{ATL-DAQ-PUB-2018-002}} (\bibinfo
  {year} {2018}{\natexlab{e}})\BibitemShut {NoStop}%
\bibitem [{\citenamefont {{ATLAS
  Collaboration}}(2019{\natexlab{b}})}]{ATL-DAQ-PUB-2019-001}%
  \BibitemOpen
  \bibfield  {author} {\bibinfo {author} {\bibnamefont {{ATLAS
  Collaboration}}},\ }\href@noop {} {\bibinfo {title} {{Trigger Menu in
  2018}}},\ \bibinfo {howpublished}
  {\href{https://cds.cern.ch/record/2693402}{ATL-DAQ-PUB-2019-001}} (\bibinfo
  {year} {2019}{\natexlab{b}})\BibitemShut {NoStop}%
\bibitem [{\citenamefont {Roche}\ \emph {et~al.}(2023)\citenamefont {Roche},
  \citenamefont {Carlson}, \citenamefont {Hayes},\ and\ \citenamefont
  {Hong}}]{dataset}%
  \BibitemOpen
  \bibfield  {author} {\bibinfo {author} {\bibfnamefont {S.}~\bibnamefont
  {Roche}}, \bibinfo {author} {\bibfnamefont {B.}~\bibnamefont {Carlson}},
  \bibinfo {author} {\bibfnamefont {C.}~\bibnamefont {Hayes}},\ and\ \bibinfo
  {author} {\bibfnamefont {T.~M.}\ \bibnamefont {Hong}},\ }\href
  {https://doi.org/10.5281/zenodo.7869418} {\bibinfo {title} {Signal and
  background samples for higgs boson decays to four b-jets}},\ \bibinfo
  {howpublished}
  {\href{https://doi.org/10.5281/zenodo.7869418}{10.5281/zenodo.7869418}}
  (\bibinfo {year} {2023})\BibitemShut {NoStop}%
\bibitem [{Far(2021)}]{Fartoukh:2790409}%
  \BibitemOpen
  \href@noop {} {\bibinfo {title} {{LHC Configuration and Operational Scenario
  for Run 3}}},\ \bibinfo {howpublished}
  {\href{https://cds.cern.ch/record/2790409}{CERN-ACC-2021-0007}} (\bibinfo
  {year} {2021})\BibitemShut {NoStop}%
\bibitem [{\citenamefont {Alwall}\ \emph {et~al.}(2014)\citenamefont {Alwall}
  \emph {et~al.}}]{Alwall:2014hca}%
  \BibitemOpen
  \bibfield  {author} {\bibinfo {author} {\bibfnamefont {J.}~\bibnamefont
  {Alwall}} \emph {et~al.},\ }\bibfield  {title} {\bibinfo {title} {{The
  automated computation of tree-level and next-to-leading order differential
  cross sections, and their matching to parton shower simulations}},\ }\href
  {https://doi.org/10.1007/JHEP07(2014)079} {\bibfield  {journal} {\bibinfo
  {journal} {J. High Energy Phys.}\ }\textbf {\bibinfo {volume} {07}}\bibinfo
  {number} { (2014)},\ \bibinfo {pages} {079}}\BibitemShut {NoStop}%
\bibitem [{\citenamefont {Sj\"ostrand}\ \emph {et~al.}(2015)\citenamefont
  {Sj\"ostrand} \emph {et~al.}}]{Sjostrand:2014zea}%
  \BibitemOpen
\bibfield  {number} {  }\bibfield  {author} {\bibinfo {author} {\bibfnamefont
  {T.}~\bibnamefont {Sj\"ostrand}} \emph {et~al.},\ }\bibfield  {title}
  {\bibinfo {title} {{An introduction to PYTHIA 8.2}},\ }\href
  {https://doi.org/10.1016/j.cpc.2015.01.024} {\bibfield  {journal} {\bibinfo
  {journal} {Comput. Phys. Commun.}\ }\textbf {\bibinfo {volume} {191}},\
  \bibinfo {pages} {159} (\bibinfo {year} {2015})}\BibitemShut {NoStop}%
\bibitem [{\citenamefont {Bierlich}\ \emph {et~al.}(2022)\citenamefont
  {Bierlich} \emph {et~al.}}]{Bierlich:2022pfr}%
  \BibitemOpen
  \bibfield  {author} {\bibinfo {author} {\bibfnamefont {C.}~\bibnamefont
  {Bierlich}} \emph {et~al.},\ }\bibfield  {title} {\bibinfo {title} {{A
  comprehensive guide to the physics and usage of PYTHIA 8.3}},\ }\bibfield
  {journal} {\bibinfo  {journal} {SciPost Phys. Codebases}\ }\textbf {\bibinfo
  {volume} {8}},\ \href {https://doi.org/10.21468/SciPostPhysCodeb.8}
  {10.21468/SciPostPhysCodeb.8} (\bibinfo {year} {2022})\BibitemShut {NoStop}%
\bibitem [{\citenamefont {{ATLAS Collaboration}}(2014)}]{ATLAS:2014alx}%
  \BibitemOpen
  \bibfield  {author} {\bibinfo {author} {\bibnamefont {{ATLAS
  Collaboration}}},\ }\bibfield  {title} {\bibinfo {title} {{Measurement of the
  $Z/\gamma^*$ boson transverse momentum distribution in $pp$ collisions at
  $\sqrt{s}$ = 7 TeV with the ATLAS detector}},\ }\href
  {https://doi.org/10.1007/JHEP09(2014)145} {\bibfield  {journal} {\bibinfo
  {journal} {J. High Energy Phys.}\ }\textbf {\bibinfo {volume} {09}}\bibinfo
  {number} { (2014)},\ \bibinfo {pages} {145}}\BibitemShut {NoStop}%
\bibitem [{\citenamefont {de~Florian}\ \emph {et~al.}(2016)\citenamefont
  {de~Florian} \emph {et~al.}}]{LHCxswg:2016ypw}%
  \BibitemOpen
\bibfield  {number} {  }\bibfield  {author} {\bibinfo {author} {\bibfnamefont
  {D.}~\bibnamefont {de~Florian}} \emph {et~al.} (\bibinfo {collaboration} {LHC
  Higgs Cross Section Working Group}),\ }\href
  {https://doi.org/10.23731/CYRM-2017-002} {\bibinfo {title} {{Handbook of LHC
  Higgs Cross Sections: 4. Deciphering the Nature of the Higgs Sector}}}
  (\bibinfo {year} {2016}),\ \Eprint {https://arxiv.org/abs/1610.07922}
  {arXiv:1610.07922 [hep-ph]} \BibitemShut {NoStop}%
\bibitem [{\citenamefont {Dreyer}\ and\ \citenamefont
  {Karlberg}(2019)}]{Dreyer:2018rfu}%
  \BibitemOpen
  \bibfield  {author} {\bibinfo {author} {\bibfnamefont {F.~A.}\ \bibnamefont
  {Dreyer}}\ and\ \bibinfo {author} {\bibfnamefont {A.}~\bibnamefont
  {Karlberg}},\ }\bibfield  {title} {\bibinfo {title} {{Fully differential
  Vector-Boson Fusion Higgs Pair Production at Next-to-Next-to-Leading
  Order}},\ }\href {https://doi.org/10.1103/PhysRevD.99.074028} {\bibfield
  {journal} {\bibinfo  {journal} {Phys. Rev. D}\ }\textbf {\bibinfo {volume}
  {99}},\ \bibinfo {pages} {074028} (\bibinfo {year} {2019})}\BibitemShut
  {NoStop}%
\bibitem [{\citenamefont {Denner}\ \emph {et~al.}(2015)\citenamefont {Denner},
  \citenamefont {Dittmaier}, \citenamefont {Kallweit},\ and\ \citenamefont
  {M\"uck}}]{Denner:2014cla}%
  \BibitemOpen
  \bibfield  {author} {\bibinfo {author} {\bibfnamefont {A.}~\bibnamefont
  {Denner}}, \bibinfo {author} {\bibfnamefont {S.}~\bibnamefont {Dittmaier}},
  \bibinfo {author} {\bibfnamefont {S.}~\bibnamefont {Kallweit}},\ and\
  \bibinfo {author} {\bibfnamefont {A.}~\bibnamefont {M\"uck}},\ }\bibfield
  {title} {\bibinfo {title} {{HAWK 2.0: A Monte Carlo program for Higgs
  production in vector-boson fusion and Higgs strahlung at hadron colliders}},\
  }\href {https://doi.org/10.1016/j.cpc.2015.04.021} {\bibfield  {journal}
  {\bibinfo  {journal} {Comput. Phys. Commun.}\ }\textbf {\bibinfo {volume}
  {195}},\ \bibinfo {pages} {161} (\bibinfo {year} {2015})}\BibitemShut
  {NoStop}%
\bibitem [{\citenamefont {Bellm}\ \emph {et~al.}(2016)\citenamefont {Bellm}
  \emph {et~al.}}]{Bellm:2015jjp}%
  \BibitemOpen
  \bibfield  {author} {\bibinfo {author} {\bibfnamefont {J.}~\bibnamefont
  {Bellm}} \emph {et~al.},\ }\bibfield  {title} {\bibinfo {title} {{Herwig
  7.0/Herwig++ 3.0 release note}},\ }\href
  {https://doi.org/10.1140/epjc/s10052-016-4018-8} {\bibfield  {journal}
  {\bibinfo  {journal} {Eur. Phys. J. C}\ }\textbf {\bibinfo {volume} {76}},\
  \bibinfo {pages} {196} (\bibinfo {year} {2016})}\BibitemShut {NoStop}%
\bibitem [{\citenamefont {Bahr}\ \emph {et~al.}(2008)\citenamefont {Bahr} \emph
  {et~al.}}]{Bahr:2008pv}%
  \BibitemOpen
  \bibfield  {author} {\bibinfo {author} {\bibfnamefont {M.}~\bibnamefont
  {Bahr}} \emph {et~al.},\ }\bibfield  {title} {\bibinfo {title} {{Herwig++
  Physics and Manual}},\ }\href
  {https://doi.org/10.1140/epjc/s10052-008-0798-9} {\bibfield  {journal}
  {\bibinfo  {journal} {Eur. Phys. J. C}\ }\textbf {\bibinfo {volume} {58}},\
  \bibinfo {pages} {639} (\bibinfo {year} {2008})}\BibitemShut {NoStop}%
\bibitem [{\citenamefont {Alwall}\ \emph {et~al.}(2011)\citenamefont {Alwall},
  \citenamefont {Herquet}, \citenamefont {Maltoni}, \citenamefont {Mattelaer},\
  and\ \citenamefont {Stelzer}}]{Alwall:2011uj}%
  \BibitemOpen
  \bibfield  {author} {\bibinfo {author} {\bibfnamefont {J.}~\bibnamefont
  {Alwall}}, \bibinfo {author} {\bibfnamefont {M.}~\bibnamefont {Herquet}},
  \bibinfo {author} {\bibfnamefont {F.}~\bibnamefont {Maltoni}}, \bibinfo
  {author} {\bibfnamefont {O.}~\bibnamefont {Mattelaer}},\ and\ \bibinfo
  {author} {\bibfnamefont {T.}~\bibnamefont {Stelzer}},\ }\bibfield  {title}
  {\bibinfo {title} {{MadGraph 5 : Going Beyond}},\ }\href
  {https://doi.org/10.1007/JHEP06(2011)128} {\bibfield  {journal} {\bibinfo
  {journal} {J. High Energy Phys.}\ }\textbf {\bibinfo {volume} {06}}\bibinfo
  {number} { (2011)},\ \bibinfo {pages} {128}}\BibitemShut {NoStop}%
\bibitem [{\citenamefont {Nackenhorst}(2015)}]{Nackenhorst:2015yjt}%
  \BibitemOpen
\bibfield  {number} {  }\bibfield  {author} {\bibinfo {author} {\bibfnamefont
  {O.}~\bibnamefont {Nackenhorst}},\ }\href@noop {} {\bibinfo {title} {{Search
  for the Standard Model Higgs boson produced in association with $t\bar{t}$
  and decaying into $b\bar{b}$ at $\sqrt{s} = 8$ TeV with the ATLAS detector
  using the Matrix Element Method}}},\ \bibinfo {howpublished}
  {\href{http://cds.cern.ch/record/2063972}{CERN-THESIS-2015-186}} (\bibinfo
  {year} {2015})\BibitemShut {NoStop}%
\bibitem [{\citenamefont {Connelly}(2016)}]{Connelly:2016nmt}%
  \BibitemOpen
  \bibfield  {author} {\bibinfo {author} {\bibfnamefont {I.~A.}\ \bibnamefont
  {Connelly}},\ }\href@noop {} {\bibinfo {title} {{A search for the Higgs boson
  decaying to two b-quarks in association with a dileptonically decaying
  top-quark pair with the ATLAS detector}}},\ \bibinfo {howpublished}
  {\href{http://cds.cern.ch/record/2271523}{Ph.D. thesis}} (\bibinfo {year}
  {2016})\BibitemShut {NoStop}%
\bibitem [{\citenamefont {{CMS
  Collaboration}}(2013{\natexlab{a}})}]{CMS:2012feb}%
  \BibitemOpen
  \bibfield  {author} {\bibinfo {author} {\bibnamefont {{CMS Collaboration}}},\
  }\bibfield  {title} {\bibinfo {title} {{Identification of b-Quark Jets with
  the CMS Experiment}},\ }\href {https://doi.org/10.1088/1748-0221/8/04/P04013}
  {\bibfield  {journal} {\bibinfo  {journal} {J. Instrum.}\ }\textbf {\bibinfo
  {volume} {8}},\ \bibinfo {pages} {P04013}}\BibitemShut {NoStop}%
\bibitem [{\citenamefont {de~Favereau}\ \emph {et~al.}(2014)\citenamefont
  {de~Favereau}, \citenamefont {Delaere}, \citenamefont {Demin}, \citenamefont
  {Giammanco}, \citenamefont {Lema\^\i{}tre}, \citenamefont {Mertens},\ and\
  \citenamefont {Selvaggi}}]{deFavereau:2013fsa}%
  \BibitemOpen
  \bibfield  {author} {\bibinfo {author} {\bibfnamefont {J.}~\bibnamefont
  {de~Favereau}}, \bibinfo {author} {\bibfnamefont {C.}~\bibnamefont
  {Delaere}}, \bibinfo {author} {\bibfnamefont {P.}~\bibnamefont {Demin}},
  \bibinfo {author} {\bibfnamefont {A.}~\bibnamefont {Giammanco}}, \bibinfo
  {author} {\bibfnamefont {V.}~\bibnamefont {Lema\^\i{}tre}}, \bibinfo {author}
  {\bibfnamefont {A.}~\bibnamefont {Mertens}},\ and\ \bibinfo {author}
  {\bibfnamefont {M.}~\bibnamefont {Selvaggi}} (\bibinfo {collaboration}
  {DELPHES 3}),\ }\bibfield  {title} {\bibinfo {title} {{DELPHES 3, A modular
  framework for fast simulation of a generic collider experiment}},\ }\href
  {https://doi.org/10.1007/JHEP02(2014)057} {\bibfield  {journal} {\bibinfo
  {journal} {J. High Energy Phys.}\ }\textbf {\bibinfo {volume} {02}}\bibinfo
  {number} { (2014)},\ \bibinfo {pages} {057}}\BibitemShut {NoStop}%
\bibitem [{\citenamefont {Ovyn}\ \emph {et~al.}(2009)\citenamefont {Ovyn},
  \citenamefont {Rouby},\ and\ \citenamefont {Lemaitre}}]{Ovyn:2009tx}%
  \BibitemOpen
\bibfield  {number} {  }\bibfield  {author} {\bibinfo {author} {\bibfnamefont
  {S.}~\bibnamefont {Ovyn}}, \bibinfo {author} {\bibfnamefont {X.}~\bibnamefont
  {Rouby}},\ and\ \bibinfo {author} {\bibfnamefont {V.}~\bibnamefont
  {Lemaitre}},\ }\href@noop {} {\bibinfo {title} {{DELPHES, a framework for
  fast simulation of a generic collider experiment}}} (\bibinfo {year}
  {2009}),\ \Eprint {https://arxiv.org/abs/0903.2225} {arXiv:0903.2225
  [hep-ph]} \BibitemShut {NoStop}%
\bibitem [{\citenamefont {Cacciari}\ \emph {et~al.}(2012)\citenamefont
  {Cacciari}, \citenamefont {Salam},\ and\ \citenamefont
  {Soyez}}]{Cacciari:2011ma}%
  \BibitemOpen
  \bibfield  {author} {\bibinfo {author} {\bibfnamefont {M.}~\bibnamefont
  {Cacciari}}, \bibinfo {author} {\bibfnamefont {G.~P.}\ \bibnamefont
  {Salam}},\ and\ \bibinfo {author} {\bibfnamefont {G.}~\bibnamefont {Soyez}},\
  }\bibfield  {title} {\bibinfo {title} {{FastJet User Manual}},\ }\href
  {https://doi.org/10.1140/epjc/s10052-012-1896-2} {\bibfield  {journal}
  {\bibinfo  {journal} {Eur. Phys. J. C}\ }\textbf {\bibinfo {volume} {72}},\
  \bibinfo {pages} {1896} (\bibinfo {year} {2012})}\BibitemShut {NoStop}%
\bibitem [{\citenamefont {Cacciari}\ \emph {et~al.}(2008)\citenamefont
  {Cacciari}, \citenamefont {Salam},\ and\ \citenamefont
  {Soyez}}]{Cacciari:2008gp}%
  \BibitemOpen
  \bibfield  {author} {\bibinfo {author} {\bibfnamefont {M.}~\bibnamefont
  {Cacciari}}, \bibinfo {author} {\bibfnamefont {G.~P.}\ \bibnamefont
  {Salam}},\ and\ \bibinfo {author} {\bibfnamefont {G.}~\bibnamefont {Soyez}},\
  }\bibfield  {title} {\bibinfo {title} {{The anti-$k_t$ jet clustering
  algorithm}},\ }\href {https://doi.org/10.1088/1126-6708/2008/04/063}
  {\bibfield  {journal} {\bibinfo  {journal} {J. High Energy Phys.}\ }\textbf
  {\bibinfo {volume} {04}}\bibinfo  {number} { (2008)},\ \bibinfo {pages}
  {063}}\BibitemShut {NoStop}%
\bibitem [{\citenamefont {{ATLAS
  Collaboration}}(2021{\natexlab{d}})}]{ATLAS:2020cli}%
  \BibitemOpen
\bibfield  {number} {  }\bibfield  {author} {\bibinfo {author} {\bibnamefont
  {{ATLAS Collaboration}}},\ }\bibfield  {title} {\bibinfo {title} {{Jet energy
  scale and resolution measured in proton\textendash{}proton collisions at
  $\sqrt{s}=13$~TeV with the ATLAS detector}},\ }\href
  {https://doi.org/10.1140/epjc/s10052-021-09402-3} {\bibfield  {journal}
  {\bibinfo  {journal} {Eur. Phys. J. C}\ }\textbf {\bibinfo {volume} {81}},\
  \bibinfo {pages} {689} (\bibinfo {year} {2021}{\natexlab{d}})}\BibitemShut
  {NoStop}%
\bibitem [{\citenamefont {{CMS Collaboration}}(2021)}]{CMS-DP-2021-033}%
  \BibitemOpen
  \bibfield  {author} {\bibinfo {author} {\bibnamefont {{CMS Collaboration}}},\
  }\href@noop {} {\bibinfo {title} {{Jet energy scale and resolution
  measurement with Run~2 Legacy Data Collected by CMS at 13~TeV}}},\ \bibinfo
  {howpublished} {\href{http://cds.cern.ch/record/2792322}{CMS-DP-2021-033}}
  (\bibinfo {year} {2021})\BibitemShut {NoStop}%
\bibitem [{\citenamefont {{CMS Collaboration}}(2013{\natexlab{b}})}]{CMSBTag}%
  \BibitemOpen
  \bibfield  {author} {\bibinfo {author} {\bibnamefont {{CMS Collaboration}}},\
  }\bibfield  {title} {\bibinfo {title} {Identification of b-quark jets with
  the {CMS} experiment},\ }\href
  {https://doi.org/10.1088/1748-0221/8/04/p04013} {\bibfield  {journal}
  {\bibinfo  {journal} {J. Instrum.}\ }\textbf {\bibinfo {volume} {8}}\bibinfo
  {number} { (04)},\ \bibinfo {pages} {P04013}}\BibitemShut {NoStop}%
\bibitem [{ATL(2022)}]{ATLAS:2022rkn}%
  \BibitemOpen
\bibfield  {number} {  }\href@noop {} {\bibinfo {title} {{Graph Neural Network
  Jet Flavour Tagging with the ATLAS Detector}}},\ \bibinfo {howpublished}
  {\href{http://cds.cern.ch/record/2811135}{ATL-PHYS-PUB-2022-027}} (\bibinfo
  {year} {2022})\BibitemShut {NoStop}%
\bibitem [{\citenamefont {Bols}\ \emph {et~al.}(2020)\citenamefont {Bols},
  \citenamefont {Kieseler}, \citenamefont {Verzetti}, \citenamefont {Stoye},\
  and\ \citenamefont {Stakia}}]{Bols:2020bkb}%
  \BibitemOpen
  \bibfield  {author} {\bibinfo {author} {\bibfnamefont {E.}~\bibnamefont
  {Bols}}, \bibinfo {author} {\bibfnamefont {J.}~\bibnamefont {Kieseler}},
  \bibinfo {author} {\bibfnamefont {M.}~\bibnamefont {Verzetti}}, \bibinfo
  {author} {\bibfnamefont {M.}~\bibnamefont {Stoye}},\ and\ \bibinfo {author}
  {\bibfnamefont {A.}~\bibnamefont {Stakia}},\ }\bibfield  {title} {\bibinfo
  {title} {{Jet Flavour Classification Using DeepJet}},\ }\href
  {https://doi.org/10.1088/1748-0221/15/12/P12012} {\bibfield  {journal}
  {\bibinfo  {journal} {J. Instrum.}\ }\textbf {\bibinfo {volume} {15}}\bibinfo
   {number} { (2020)},\ \bibinfo {pages} {P12012}}\BibitemShut {NoStop}%
\bibitem [{\citenamefont {Demin}\ and\ \citenamefont
  {Selvaggi}(2017)}]{delphesPhotonEfficiency}%
  \BibitemOpen
\bibfield  {number} {  }\bibfield  {author} {\bibinfo {author} {\bibfnamefont
  {P.}~\bibnamefont {Demin}}\ and\ \bibinfo {author} {\bibfnamefont
  {M.}~\bibnamefont {Selvaggi}},\ }\href@noop {} {\bibinfo {title} {{Delphes
  card: CMS}}},\ \bibinfo {howpublished} {GitHub,
  \href{https://github.com/delphes/delphes/blob/7b7cc48d2c23d91a4377bbe83953fb58ae7aa317/cards/delphes_card_CMS.tcl\#L464}{cards/delphes\_card\_CMS.tcl\#L464}
  in delphes/delphes/ blob/7b7cc48d2c23d91a4377bbe83953fb58ae7aa317/} (\bibinfo
  {year} {2017})\BibitemShut {NoStop}%
\bibitem [{\citenamefont {{ATLAS
  Collaboration}}(2012{\natexlab{b}})}]{ATLAS:2012nks}%
  \BibitemOpen
  \bibfield  {author} {\bibinfo {author} {\bibnamefont {{ATLAS
  Collaboration}}},\ }\bibfield  {title} {\bibinfo {title} {{Performance of the
  ATLAS Trigger System in 2010}},\ }\href
  {https://doi.org/10.1140/epjc/s10052-011-1849-1} {\bibfield  {journal}
  {\bibinfo  {journal} {Eur. Phys. J. C}\ }\textbf {\bibinfo {volume} {72}},\
  \bibinfo {pages} {1849} (\bibinfo {year} {2012}{\natexlab{b}})}\BibitemShut
  {NoStop}%
\bibitem [{\citenamefont {{ATLAS
  Collaboration}}(2017{\natexlab{b}})}]{Aaboud:2016leb}%
  \BibitemOpen
  \bibfield  {author} {\bibinfo {author} {\bibnamefont {{ATLAS
  Collaboration}}},\ }\bibfield  {title} {\bibinfo {title} {{Performance of the
  ATLAS Trigger System in 2015}},\ }\href
  {https://doi.org/10.1140/epjc/s10052-017-4852-3} {\bibfield  {journal}
  {\bibinfo  {journal} {Eur. Phys. J. C}\ }\textbf {\bibinfo {volume} {77}},\
  \bibinfo {pages} {317} (\bibinfo {year} {2017}{\natexlab{b}})}\BibitemShut
  {NoStop}%
\bibitem [{\citenamefont {{CMS Collaboration}}(2017)}]{CMS:2016ngn}%
  \BibitemOpen
  \bibfield  {author} {\bibinfo {author} {\bibnamefont {{CMS Collaboration}}},\
  }\bibfield  {title} {\bibinfo {title} {{The CMS trigger system}},\ }\href
  {https://doi.org/10.1088/1748-0221/12/01/P01020} {\bibfield  {journal}
  {\bibinfo  {journal} {J. Instrum.}\ }\textbf {\bibinfo {volume} {12}}\bibinfo
   {number} { (01)},\ \bibinfo {pages} {P01020}}\BibitemShut {NoStop}%
\bibitem [{\citenamefont {{ATLAS
  Collaboration}}(2020{\natexlab{e}})}]{ATLAS:2020osu}%
  \BibitemOpen
\bibfield  {number} {  }\bibfield  {author} {\bibinfo {author} {\bibnamefont
  {{ATLAS Collaboration}}},\ }\bibfield  {title} {\bibinfo {title}
  {{Performance of the upgraded PreProcessor of the ATLAS Level-1 Calorimeter
  Trigger}},\ }\href {https://doi.org/10.1088/1748-0221/15/11/P11016}
  {\bibfield  {journal} {\bibinfo  {journal} {J. Instrum.}\ }\textbf {\bibinfo
  {volume} {15}}\bibinfo  {number} { (11)},\ \bibinfo {pages}
  {P11016}}\BibitemShut {NoStop}%
\bibitem [{\citenamefont {Stelzer}(2011)}]{Stelzer:2011zz}%
  \BibitemOpen
\bibfield  {number} {  }\bibfield  {author} {\bibinfo {author} {\bibfnamefont
  {J.}~\bibnamefont {Stelzer}} (\bibinfo {collaboration} {ATLAS
  Collaboration}),\ }\bibfield  {title} {\bibinfo {title} {{The ATLAS high
  level trigger configuration and steering: Experience with the first 7-TeV
  collision data}},\ }\href {https://doi.org/10.1088/1742-6596/331/2/022026}
  {\bibfield  {journal} {\bibinfo  {journal} {J. Phys. Conf. Ser.}\ }\textbf
  {\bibinfo {volume} {331}},\ \bibinfo {pages} {022026} (\bibinfo {year}
  {2011})}\BibitemShut {NoStop}%
\bibitem [{\citenamefont {{CMS
  Collaboration}}(2020{\natexlab{b}})}]{Zabi:2020gjd}%
  \BibitemOpen
  \bibfield  {author} {\bibinfo {author} {\bibnamefont {{CMS Collaboration}}},\
  }\href@noop {} {\bibinfo {title} {{The Phase-2 Upgrade of the CMS Level-1
  Trigger}}},\ \bibinfo {howpublished}
  {\href{https://cds.cern.ch/record/2714892}{CMS-TDR-021}} (\bibinfo {year}
  {2020}{\natexlab{b}})\BibitemShut {NoStop}%
\bibitem [{\citenamefont {{ATLAS Collaboration}}(2013)}]{ATLAS:2013lic}%
  \BibitemOpen
  \bibfield  {author} {\bibinfo {author} {\bibnamefont {{ATLAS
  Collaboration}}},\ }\href@noop {} {\bibinfo {title} {{Technical Design Report
  for the Phase-I Upgrade of the ATLAS TDAQ System}}},\ \bibinfo {howpublished}
  {\href{https://cds.cern.ch/record/1602235/}{ATLAS-TDR-023}} (\bibinfo {year}
  {2013})\BibitemShut {NoStop}%
\bibitem [{\citenamefont {{ATLAS
  Collaboration}}(2017{\natexlab{c}})}]{ATLAS-TDR-029}%
  \BibitemOpen
  \bibfield  {author} {\bibinfo {author} {\bibnamefont {{ATLAS
  Collaboration}}},\ }\href {https://doi.org/10.17181/CERN.2LBB.4IAL} {\bibinfo
  {title} {{Technical Design Report for the Phase-II Upgrade of the ATLAS TDAQ
  System}}},\ \bibinfo {howpublished}
  {\href{http://cds.cern.ch/record/2285584}{ATLAS-TDR-029}} (\bibinfo {year}
  {2017}{\natexlab{c}})\BibitemShut {NoStop}%
\bibitem [{\citenamefont {Hoecker}\ \emph {et~al.}(2007)\citenamefont {Hoecker}
  \emph {et~al.}}]{Hocker:2007ht}%
  \BibitemOpen
  \bibfield  {author} {\bibinfo {author} {\bibfnamefont {A.}~\bibnamefont
  {Hoecker}} \emph {et~al.},\ }\href@noop {} {\bibinfo {title} {{TMVA - Toolkit
  for Multivariate Data Analysis}}} (\bibinfo {year} {2007}),\ \Eprint
  {https://arxiv.org/abs/physics/0703039} {arXiv:physics/0703039
  [physics.data-an]} \BibitemShut {NoStop}%
\bibitem [{\citenamefont {Freund}\ and\ \citenamefont
  {Schapire}(1995)}]{adaboost}%
  \BibitemOpen
  \bibfield  {author} {\bibinfo {author} {\bibfnamefont {Y.}~\bibnamefont
  {Freund}}\ and\ \bibinfo {author} {\bibfnamefont {R.~E.}\ \bibnamefont
  {Schapire}},\ }\bibfield  {title} {\bibinfo {title} {A desicion-theoretic
  generalization of on-line learning and an application to boosting},\ }in\
  \href@noop {} {\emph {\bibinfo {booktitle} {Computational Learning
  Theory}}},\ \bibinfo {editor} {edited by\ \bibinfo {editor} {\bibfnamefont
  {P.}~\bibnamefont {Vit{\'a}nyi}}}\ (\bibinfo  {publisher} {Springer Berlin
  Heidelberg},\ \bibinfo {address} {Berlin, Heidelberg},\ \bibinfo {year}
  {1995})\ pp.\ \bibinfo {pages} {23--37}\BibitemShut {NoStop}%
\bibitem [{\citenamefont {Barger}\ \emph {et~al.}(1991)\citenamefont {Barger},
  \citenamefont {Cheung}, \citenamefont {Han}, \citenamefont {Ohnemus},\ and\
  \citenamefont {Zeppenfeld}}]{Barger:1991ib}%
  \BibitemOpen
  \bibfield  {author} {\bibinfo {author} {\bibfnamefont {V.~D.}\ \bibnamefont
  {Barger}}, \bibinfo {author} {\bibfnamefont {K.-m.}\ \bibnamefont {Cheung}},
  \bibinfo {author} {\bibfnamefont {T.}~\bibnamefont {Han}}, \bibinfo {author}
  {\bibfnamefont {J.}~\bibnamefont {Ohnemus}},\ and\ \bibinfo {author}
  {\bibfnamefont {D.}~\bibnamefont {Zeppenfeld}},\ }\bibfield  {title}
  {\bibinfo {title} {{A Comparative study of the benefits of forward jet
  tagging in heavy Higgs production at the SSC}},\ }\href
  {https://doi.org/10.1103/PhysRevD.44.1426} {\bibfield  {journal} {\bibinfo
  {journal} {Phys. Rev. D}\ }\textbf {\bibinfo {volume} {44}},\ \bibinfo
  {pages} {1426} (\bibinfo {year} {1991})}\BibitemShut {NoStop}%
\bibitem [{\citenamefont {Eboli}\ and\ \citenamefont
  {Zeppenfeld}(2000)}]{Eboli:2000ze}%
  \BibitemOpen
  \bibfield  {author} {\bibinfo {author} {\bibfnamefont {O.~J.~P.}\
  \bibnamefont {Eboli}}\ and\ \bibinfo {author} {\bibfnamefont
  {D.}~\bibnamefont {Zeppenfeld}},\ }\bibfield  {title} {\bibinfo {title}
  {{Observing an invisible Higgs boson}},\ }\href
  {https://doi.org/10.1016/S0370-2693(00)01213-2} {\bibfield  {journal}
  {\bibinfo  {journal} {Phys. Lett. B}\ }\textbf {\bibinfo {volume} {495}},\
  \bibinfo {pages} {147} (\bibinfo {year} {2000})}\BibitemShut {NoStop}%
\bibitem [{\citenamefont {Curtin}\ \emph {et~al.}(2015)\citenamefont {Curtin},
  \citenamefont {Essig},\ and\ \citenamefont {Zhong}}]{Curtin:2014pda}%
  \BibitemOpen
  \bibfield  {author} {\bibinfo {author} {\bibfnamefont {D.}~\bibnamefont
  {Curtin}}, \bibinfo {author} {\bibfnamefont {R.}~\bibnamefont {Essig}},\ and\
  \bibinfo {author} {\bibfnamefont {Y.-M.}\ \bibnamefont {Zhong}},\ }\bibfield
  {title} {\bibinfo {title} {{Uncovering light scalars with exotic Higgs decays
  to $ b\overline{b}{\mu}^{+}{\mu}^{-} $}},\ }\href
  {https://doi.org/10.1007/JHEP06(2015)025} {\bibfield  {journal} {\bibinfo
  {journal} {JHEP}\ }\textbf {\bibinfo {volume} {06}},\ \bibinfo {pages}
  {025}}\BibitemShut {NoStop}%
\bibitem [{\citenamefont {{ATLAS
  Collaboration}}(2021{\natexlab{e}})}]{ATLAS:2020vbr}%
  \BibitemOpen
  \bibfield  {author} {\bibinfo {author} {\bibnamefont {{ATLAS
  Collaboration}}},\ }\bibfield  {title} {\bibinfo {title} {{Search for Higgs
  boson production in association with a high-energy photon via vector-boson
  fusion with decay into bottom quark pairs at $\sqrt{s}$=13 TeV with the ATLAS
  detector}},\ }\href {https://doi.org/10.1007/JHEP03(2021)268} {\bibfield
  {journal} {\bibinfo  {journal} {J. High Energy Phys.}\ }\textbf {\bibinfo
  {volume} {03}}\bibinfo  {number} { (2021)},\ \bibinfo {pages}
  {268}}\BibitemShut {NoStop}%
\bibitem [{\citenamefont {{ATLAS
  Collaboration}}(2022{\natexlab{g}})}]{ATLAS:2021tnq}%
  \BibitemOpen
\bibfield  {number} {  }\bibfield  {author} {\bibinfo {author} {\bibnamefont
  {{ATLAS Collaboration}}},\ }\bibfield  {title} {\bibinfo {title}
  {{Performance of the ATLAS Level-1 topological trigger in Run~2}},\ }\href
  {https://doi.org/10.1140/epjc/s10052-021-09807-0} {\bibfield  {journal}
  {\bibinfo  {journal} {Eur. Phys. J. C}\ }\textbf {\bibinfo {volume} {82}},\
  \bibinfo {pages} {7} (\bibinfo {year} {2022}{\natexlab{g}})}\BibitemShut
  {NoStop}%
\bibitem [{\citenamefont {{ATLAS
  Collaboration}}(2020{\natexlab{f}})}]{Aad:2019wsl}%
  \BibitemOpen
  \bibfield  {author} {\bibinfo {author} {\bibnamefont {{ATLAS
  Collaboration}}},\ }\bibfield  {title} {\bibinfo {title} {{Performance of
  electron and photon triggers in ATLAS during LHC Run 2}},\ }\href
  {https://doi.org/10.1140/epjc/s10052-019-7500-2} {\bibfield  {journal}
  {\bibinfo  {journal} {Eur. Phys. J. C}\ }\textbf {\bibinfo {volume} {80}},\
  \bibinfo {pages} {47} (\bibinfo {year} {2020}{\natexlab{f}})}\BibitemShut
  {NoStop}%
\bibitem [{\citenamefont {Apollinari}\ \emph {et~al.}(2015)\citenamefont
  {Apollinari}, \citenamefont {Br\"uning}, \citenamefont {Nakamoto},\ and\
  \citenamefont {Rossi}}]{Apollinari:2015wtw}%
  \BibitemOpen
  \bibfield  {author} {\bibinfo {author} {\bibfnamefont {G.}~\bibnamefont
  {Apollinari}}, \bibinfo {author} {\bibfnamefont {O.}~\bibnamefont
  {Br\"uning}}, \bibinfo {author} {\bibfnamefont {T.}~\bibnamefont
  {Nakamoto}},\ and\ \bibinfo {author} {\bibfnamefont {L.}~\bibnamefont
  {Rossi}},\ }\bibfield  {title} {\bibinfo {title} {{High Luminosity Large
  Hadron Collider HL-LHC}},\ }\href {https://doi.org/10.5170/CERN-2015-005.1}
  {\bibfield  {journal} {\bibinfo  {journal} {CERN Yellow Rep.}\ }\textbf
  {\bibinfo {volume} {5}},\ \bibinfo {pages} {1} (\bibinfo {year}
  {2015})}\BibitemShut {NoStop}%
\bibitem [{\citenamefont {Govorkova}\ \emph {et~al.}(2022)\citenamefont
  {Govorkova} \emph {et~al.}}]{Govorkova:2021utb}%
  \BibitemOpen
  \bibfield  {author} {\bibinfo {author} {\bibfnamefont {E.}~\bibnamefont
  {Govorkova}} \emph {et~al.},\ }\bibfield  {title} {\bibinfo {title}
  {{Autoencoders on field-programmable gate arrays for real-time, unsupervised
  new physics detection at 40 MHz at the Large Hadron Collider}},\ }\href
  {https://doi.org/10.1038/s42256-022-00441-3} {\bibfield  {journal} {\bibinfo
  {journal} {Nature Mach. Intell.}\ }\textbf {\bibinfo {volume} {4}},\ \bibinfo
  {pages} {154} (\bibinfo {year} {2022})}\BibitemShut {NoStop}%
\bibitem [{\citenamefont {Roche}\ \emph {et~al.}(2024)\citenamefont {Roche},
  \citenamefont {Bayer}, \citenamefont {Carlson}, \citenamefont {Ouligian},
  \citenamefont {Serhiayenka}, \citenamefont {Stelzer},\ and\ \citenamefont
  {Hong}}]{Roche:2023int}%
  \BibitemOpen
  \bibfield  {author} {\bibinfo {author} {\bibfnamefont {S.}~\bibnamefont
  {Roche}}, \bibinfo {author} {\bibfnamefont {Q.}~\bibnamefont {Bayer}},
  \bibinfo {author} {\bibfnamefont {B.}~\bibnamefont {Carlson}}, \bibinfo
  {author} {\bibfnamefont {W.}~\bibnamefont {Ouligian}}, \bibinfo {author}
  {\bibfnamefont {P.}~\bibnamefont {Serhiayenka}}, \bibinfo {author}
  {\bibfnamefont {J.}~\bibnamefont {Stelzer}},\ and\ \bibinfo {author}
  {\bibfnamefont {T.~M.}\ \bibnamefont {Hong}},\ }\bibfield  {title} {\bibinfo
  {title} {{Nanosecond anomaly detection with decision trees and real-time
  application to exotic Higgs decays}},\ }\href
  {https://doi.org/10.1038/s41467-024-47704-8} {\bibfield  {journal} {\bibinfo
  {journal} {Nature Commun.}\ }\textbf {\bibinfo {volume} {15}},\ \bibinfo
  {pages} {3527} (\bibinfo {year} {2024})}\BibitemShut {NoStop}%
\end{thebibliography}%

\end{document}